\global\def\draftcontrol{0}

   \def\versionno{ n=2thermo -- draft   }
\catcode`\@=11

\expandafter\ifx\csname draftcontrol\endcsname\relax\global\def\draftcontrol{0}
\fi

{\count255=\time\divide\count255 by 60
\xdef\hourmin{\number\count255}
\multiply\count255 by-60\advance\count255 by\time
\xdef\hourmin{\hourmin:\ifnum\count255<10 0\fi\the\count255}}
\def\draftdate{\number\month/\number\day/\number\year\ \ \ \hourmin }

\newcommand\makepapertitle{\par
  \begingroup
    \renewcommand\thefootnote{\@fnsymbol\c@footnote}%
    \def\@makefnmark{\rlap{\@textsuperscript{\normalfont\@thefnmark}}}%
    \long\def\@makefntext##1{\parindent 1em\noindent
            \hb@xt@1.8em{%
                \hss\@textsuperscript{\normalfont\@thefnmark}}##1}%
     \newpage
     \global\@topnum\z@   % Prevents figures from going at top of page.
     \@makepapertitle
     \thispagestyle{empty}\@thanks
  \endgroup
  \setcounter{footnote}{0}%
  \global\let\thanks\relax
  \global\let\makepapertitle\relax
  \global\let\@makepapertitle\relax
  \global\let\@thanks\@empty
  \global\let\@author\@empty
  \global\let\@date\@empty
  \global\let\@title\@empty
  \global\let\title\relax
  \global\let\author\relax
  \global\let\date\relax
  \global\let\and\relax
  \def\version{\let\version\@version\@gobble}
}
\def\@makepapertitle{%
  \newpage
   \ifnum\draftcontrol=1 {}
   \version\versionno
   \vskip 3em%
   \else
   \hfill\hbox to 3cm {\parbox{4cm}{\@pubnum}\hss}%
   \vskip 3em%
   \fi
   \begin{center}%
   \let \footnote \thanks
     {\LARGE {\@title}}%
     \vskip 1.5em%
     {\normalsize%\large
       \lineskip .5em%
       \begin{tabular}[t]{c}%
         \@author
       \end{tabular}\par}%
     \vskip 1.5em%
     {\@bstract}%
     \end{center}%
     \vskip 1.5em
     \@date%
   \par
}

\gdef\@pubnum{}
\def\pubnum#1{%
  \gdef\@pubnum{#1}}

\gdef\@bstract{}
\def\Abstract#1{%
  \gdef\@bstract{%
   \parbox{\textwidth-0pc}{%
   \centerline{\bf Abstract}\penalty1000%
\kern.2cm%
\noindent%\abstractfont \baselineskip=12pt
\renewcommand\baselinestretch{1.0}%
{#1}}}
}

\def\ps@paper{\let\@mkboth\@gobbletwo%
     \ifnum\draftcontrol=1
    \def\@oddfoot{\hbox to \textwidth{\tiny \versionno \hfil\tiny\draftdate}%
    \hskip -\textwidth \hbox to \textwidth{\hfil\rm\thepage\hfil}}%
     \else\def\@oddfoot{\hbox to \textwidth{\hfil\rm\thepage\hfil}}
     \fi
     \let\@evenfoot\@oddfoot
}
\def\body{\clearpage
          \pagestyle{paper}
    }
\def\@version#1{\ifnum\draftcontrol=1
\typeout{}\typeout{#1}\typeout{}
\vskip3mm\centerline{\hbox{\fbox{\normalsize{\tt DRAFT -- #1 -- }
                   {\draftdate}}}}\vskip3mm
\fi}
\let\version\@version
\long\def\eqlabel#1{\ifnum\draftcontrol=1
                    \tag@false  % there are some problems with multline without this
                    \tag*{(\theequation) \hbox to -0.2cm{\hspace{0cm}\small{#1}\hss}}
                    \refstepcounter{equation}
                    \edef\@currentlabel{\theequation}
                    \ltx@label{#1}          % use old LaTeX \label instead of new definition
                    \else
                    \label{#1}
                    \fi
                    }
\let\st@bibitem\@bibitem
\let\st@lbibitem\@lbibitem
\ifnum\draftcontrol=1
  \def\@bibitem#1{%
    \st@bibitem{#1}\a@@label{#1}\ignorespaces}
  \def\@lbibitem[#1]#2{%
    \st@lbibitem[#1]{#2}\a@@label{#2}\ignorespaces}
  \def\a@@label#1{%
    \gdef\a@lab{\smash{\normalfont\small#1}}
    \ifvmode
      \if@inlabel
        \global\setbox\@labels\hbox{%
          \llap{\a@lab\let\a@lab\relax
                \kern\@totalleftmargin\kern\marginparsep}%
          \box\@labels}%
      \fi
    \fi}
\fi
\documentclass[12pt,letterpaper]{article}
\usepackage{amsmath,amssymb,array,calc,epsfig}
\usepackage{psfrag,verbatim,bm}
\usepackage[nosort]{cite}
\ifnum\draftcontrol=1
\fi
\renewcommand\baselinestretch{1.25}
\renewcommand\section{\@startsection {section}{1}{\z@}%
                                   {-3.5ex \@plus -1ex \@minus -.2ex}%
                                   {2.3ex \@plus.2ex}%
                                   {\normalfont\large\bfseries}}
\renewcommand\subsection{\@startsection{subsection}{2}{\z@}%
                                   {-3.25ex\@plus -1ex \@minus -.2ex}%
                                   {1.5ex \@plus .2ex}%
                                   {\normalfont\normalsize\bfseries}}
\renewcommand\subsubsection{\@startsection{subsubsection}{3}{\z@}%
                                   {-3.25ex\@plus -1ex \@minus -.2ex}%
                                   {1.5ex \@plus .2ex}%
                                   {\normalfont\normalsize\it}}
\renewcommand\paragraph{\@startsection{paragraph}{4}{\z@}%
                                   {-3.25ex\@plus -1ex \@minus -.2ex}%
                                   {1.5ex \@plus .2ex}%
                                   {\normalfont\normalsize\bf}}

\numberwithin{equation}{section}

\def\ie{{\it i.e.}}

\def\revise#1       {\raisebox{-0em}{\rule{3pt}{1em}}%
                     \marginpar{\raisebox{.5em}{\vrule width3pt\
                     \vrule width0pt height 0pt depth0.5em
                     \hbox to 0cm{\hspace{0cm}{%
                     \parbox[t]{4em}{\raggedright\footnotesize{#1}}}\hss}}}}

\newcommand\nxt[1]  {\\\fnxt#1}

\def\cale         {{\cal E}}
\def\calf         {{\cal F}}

\def\call         {{\cal L}}
\def\calm         {{\cal M}}
\def\caln         {{\cal N}}
\def\calo         {{\cal O}}
\def\calp         {{\cal P}}

\def\reals        {{\mathbb R}}

\def\del          {\partial}

\def\ee           {{\rm e}}

\def\tr           {\mathop{\rm Tr}}

\def\sqr#1#2{{\vcenter{\vbox{\hrule height.#2pt
 \hbox{\vrule width.#2pt height#1pt \kern#1pt
 \vrule width.#2pt}\hrule height.#2pt}}}}

\newcommand{\fft}[2]{{\frac{#1}{#2}}}
\newcommand{\ft}[2]{{\textstyle{\frac{#1}{#2}}}}
\def\jsquare{\mathop{\mathchoice{\sqr{8}{32}}{\sqr{8}{32}}
{\sqr{6.3}{9}}{\sqr{4.5}{9}}}}

\def\a{\alpha}

\def\r{\rho}
\def\dd{\delta}
\def\hdd{\hat{\delta}}

\def\c{\chi}
\def\ga{\gamma}
\def\ee{\epsilon}
\def\hx{\hat{x}}

\def\xh{\hat{x}}

\def\aa1{\phi}
\def\cc1{\psi}

\def\hr{\hat{\rho}}

\def\hdd{\hat{\delta}}
\def\tQ{\tilde{Q}}
\def\k{\kappa}
\catcode`\@=12

\begin{document}

%%%
%%%%%% text starts here
%%%%%%%%%

\title{Thermodynamics of the $\caln=2^*$ strongly coupled plasma}

\pubnum{%
UWO-TH-07/01
MCTP-07-04}
\date{January 2007}

\author{
Alex Buchel$ ^{1,2}$,  Stan Deakin$ ^1$, Patrick Kerner$^{1}$ and James T. Liu$^{3}$\\[0.4cm]
\it $ ^1$Department of Applied Mathematics\\
\it University of Western Ontario\\
\it London, Ontario N6A 5B7, Canada\\[0.2cm]
\it $ ^2$Perimeter Institute for Theoretical Physics\\
\it Waterloo, Ontario N2J 2W9, Canada\\[0.2cm]
\it $^3$Michigan Center for Theoretical Physics\\
\it Randall Laboratory of Physics, The University of Michigan\\
\it Ann Arbor, MI 48109--1040, USA
}

\Abstract{
Gauge/string duality is a potentially important framework for
addressing the properties of the strongly coupled quark gluon plasma
produced at RHIC.  However, constructing an actual string theory dual
to QCD has so far proven elusive.  In this paper, we take a partial
step towards exploring the QCD plasma by investigating the
thermodynamics of a non-conformal system, namely the $\caln=2^*$
theory, which is obtained as a mass deformation of the conformal
$\caln=4$ gauge theory.  We find that at temperatures of order the
mass scale, the thermodynamics of the mass deformed plasma is
surprisingly close to that of the conformal gauge theory plasma.  This
suggests that many properties of the quark gluon plasma at RHIC may in
fact be well described by even relatively simple models such as that
of the conformal $\caln=4$ plasma.
}

\makepapertitle

\body

\version\versionno

%%%%%%%%%%%%%%%%%%%%%%%%%%%%%%%%%%%%%%%%
\section{Introduction }

The purpose of this paper is to explore properties of strongly coupled
nonconformal gauge theory plasma within the framework of the gauge
theory/string theory correspondence of Maldacena \cite{m1,m2}.  We are
primarily motivated by recent attempts (see \cite{raj} and references therein)
to describe the quark gluon plasma (QGP) produced at RHIC
\cite{rhic1,rhic2,rhic3,rhic4} from the dual holographic perspective.
Unfortunately, we do not as yet have a controllable string theory dual
to QCD. In fact, most applications of the Maldacena correspondence to RHIC
physics are discussed in the context of conformal $\caln=4$ supersymmetric
$SU(N)$ Yang-Mills (SYM) theory in the 't~Hooft (planar) limit and for
large 't~Hooft coupling.  Remarkably, $\caln=4$ SYM plasma
\cite{ads1,ads2,ads3,ads4,ads5,ads6,ads7,ads8,ads9} as a model for the
RHIC QGP appears to be reasonably good \cite{shu2}. We would like to
stress, however, that such an agreement is rather paradoxical.
One requires a strong 't~Hooft coupling in order to have a controllable
string theory dual, and the QGP does indeed appear to be strongly coupled.
However, it is strongly coupled because it is produced at temperatures of
order the QCD strong coupling scale, where, naively, conformal invariance
is badly broken.  So, why then does the conformal gauge theory 
plasma serve as {\it the model} for RHIC physics? 

As a first step towards answering the latter question, we study the
thermodynamics of the $\caln=2^*$ ({\it i.e.}~mass deformed $\caln=4$)
SYM plasma over a wide range of temperatures and for different mass 
deformations%
\footnote{Although there have been previous attempts to study the
thermodynamics of strongly coupled nonconformal four dimensional gauge
theory plasmas, they have been limited to the high temperature regime,
where the theory is almost conformal \cite{kt1,kt2,kt3,bl,aby}. A notable difference is 
a recent study of thermodynamics of strongly coupled 
$\caln=2$ gauge theory plasma with massive fundamental hypermultiplets \cite{rob}.}.
The expectation here is that the deformation mass scale would provide
a model for the QCD strong coupling scale, and the thermodynamics of 
the mass deformed plasma at temperatures of order the mass scale would
then serve as a more realistic model for the RHIC QGP.

The paper is organized as follows. In the next section we review the
non-extremal $\caln=2^*$ geometry constructed in \cite{bl}, the holographic
renormalization of this theory \cite{hr} and also recall the explicit
expressions for the free energy density, energy density and the entropy
density \cite{hr,bbs}.  In section~\ref{sec:gtvsp} we discuss the map
between finite temperature $\caln=2^*$ gauge theory parameters and the
parameters of the dual non-extremal geometry \cite{bpp,bl}. In
section~\ref{sec:thtpwf} we review the high-temperature thermodynamics
of the finite temperature PW flow \cite{bl,hr,bbs}.  After this, we
turn towards a numerical investigation of the flow away from the
high-temperature regime. We describe our numerical procedure in
section~\ref{sec:np}, and present the results of the analysis in
section~\ref{sec:r}.  Finally, we conclude and also outline open
problems in section~\ref{sec:cfd}.

%%%%%%%%%%%%%%%%%%%%%%%%%%%%%%%%%%%%%%%%
\section{Non-extremal $\caln=2^*$ geometry}
\label{sec:neg}

The supergravity background dual to finite
temperature $\caln=2^*$ gauge theory \cite{bl} is a deformation
of the original $AdS_5\times S^5$ geometry induced by a pair of
scalars $\alpha$ and $\chi$ of the five-dimensional gauge
supergravity. (At zero temperature, such a deformation was
constructed by Pilch and Warner \cite{pw}.) According to the
general scenario of a holographic RG flow, the asymptotic
boundary behavior of the supergravity scalars is related to the
bosonic and fermionic mass parameters of the relevant operators
inducing the RG flow in the boundary gauge theory. Based on such a
relation, and the fact that $\alpha$ and $\chi$ have conformal dimensions
two and one, respectively, we call the supergravity 
scalar $\a$ a {\it bosonic} deformation, and the supergravity scalar $\chi$  
a {\it fermionic} deformation of the D3-brane geometry.

The action of the five-dimensional gauged supergravity including the
scalars $\alpha$ and $\chi$ is given by
\begin{equation}
\begin{split}
S=&\,
\int_{\calm_5} d\xi^5 \sqrt{-g}\ \call_5\\
=&\frac{1}{4\pi G_5}\,
\int_{\calm_5} d\xi^5 \sqrt{-g}\left[\ft14 R-3 (\del\a)^2-(\del\chi)^2-
\calp\right]\,,
\end{split}
\eqlabel{action5}
\end{equation}
where the potential%
\footnote{We set the five-dimensional gauged
supergravity coupling to one. This corresponds to setting the
radius $L$ of the five-dimensional sphere in the undeformed metric
to $2$.}
\begin{equation}
\calp=\frac{1}{16}\left[\frac 13 \left(\frac{\del W}{\del
\a}\right)^2+ \left(\frac{\del W}{\del \chi}\right)^2\right]-\frac
13 W^2\,
 \eqlabel{pp}
\end{equation}
is a function of $\alpha$ and $\chi$, and is determined by the
superpotential
\begin{equation}
W=- e^{-2\alpha} - \frac{1}{2} e^{4\alpha} \cosh(2\chi)\,.
\eqlabel{supp}
\end{equation}
In our conventions, the five-dimensional Newton's constant is
\begin{equation}
G_5\equiv \frac{G_{10}}{2^5\ {\rm vol}_{S^5}}=\frac{4\pi}{N^2}\,.
\eqlabel{g5}
\end{equation}
The action \eqref{action5} yields the Einstein equations
\begin{equation}
R_{\mu\nu}=12 \del_\mu \a \del_\nu\a+ 4 \del_\mu\chi\del_\nu\chi
+\frac43 g_{\mu\nu} \calp\,,
\eqlabel{ee}
\end{equation}
as well as the equations for the scalars
\begin{equation}
\jsquare\alpha=\fft16\fft{\del\calp}{\del\alpha}\,,\qquad
\jsquare\chi=\fft12\fft{\del\calp}{\del\chi}\,.
\eqlabel{scalar}
\end{equation}

To construct a finite-temperature version of the Pilch-Warner
flow, we choose an ansatz for the metric respecting rotational
but not the Lorentzian invariance
\begin{equation}
ds_5^2=-c_1^2(r)\ dt^2 +c_2^2(r)\ \left( d x_1^2 +  d x_2^2 + d x_3^2 \right)
+  dr^2\,.\\
\eqlabel{ab}
\end{equation}
With this ansatz, the equations of motion for the background become
\begin{equation}
\begin{split}
&\a''+\a'\ \left(\ln{c_1c_2^3}\right)'
-\frac 16 \frac{\del\calp}{\del\a}=0\,,\\
&\c''+\c'\ \left(\ln{c_1c_2^3}\right)'-\frac 12 \frac{\del\calp}{\del\c}=0\,,\\
&c_1''+c_1'\ \left(\ln{c_2^3}\right)'+\frac 43c_1  \calp=0\,,\\
&c_2''+c_2'\ \left(\ln{c_1c_2^2}\right)'+\frac 43c_2  \calp=0\,,
\end{split}
\eqlabel{beom}
\end{equation}
where the prime denotes a derivative with respect to the radial
coordinate $r$.
In addition, there is a first-order constraint
\begin{equation}
\left(\a'\right)^2+\frac 13 \left(\c'\right)^2 -\frac 13
\calp-\frac 12 (\ln c_2)'(\ln c_1 c_2)' =0\,. 
\eqlabel{backconst}
\end{equation}
It was shown in \cite{bl} that any solution to \eqref{beom} and
\eqref{backconst} can be lifted to a 
full ten-dimensional solution of type IIb supergravity. This includes 
the metric, the three- and five-form fluxes, the dilaton and the axion.
In particular, the ten-dimensional 
Einstein frame metric is given by Eq.~(4.12) in \cite{bl}.

For finite temperature flows, we find it convenient to introduce a
new radial coordinate $x$ which spans the range from the horizon to
the boundary in finite coordinate distance:
\begin{equation}
1-x(r) = \frac{c_1}{c_2}\,, \qquad x\in [0,1]\,.
\eqlabel{radgauge}
\end{equation}
With this new coordinate, the black brane's horizon is at $x=1$,
while the boundary of the asymptotically $AdS_5$ space-time is at
$x=0$%
\footnote{This $x$ coordinate is inappropriate for the extremal PW
flow, since in that case $x$ is always vanishing.  In this sense,
backgrounds with and without horizons (or equivalently physics at
finite temperature and zero temperature) may have distinctly different
characteristics.}.
The background equations of motion \eqref{beom} become
\begin{equation}
\begin{split}
&c_2''+4 c_2\ (\alpha')^2-\frac {1}{x-1} c_2'-\frac{5}{c_2} (c_2')^2
+\frac 43 c_2\ (\c')^2 =0\,,\\
&\alpha''+\frac {1}{x-1} \, \alpha'-
\frac{1}{12\, \calp c_2^2 (x-1)}\biggl[     (x-1) \left(6 (\alpha')^2+2(\chi')^2\right) c_2^2\\
&\kern16em -3 c_2' c_2-6 (c_2')^2 (x-1) \biggr] \,
 \frac{\del\calp}{\del\a} =0\,,\\
&\chi''+\frac {1}{x-1} \, \chi'-
\frac{1}{4\, \calp c_2^2 (x-1)}\biggl[     (x-1) \left(6 (\alpha')^2+2(\chi')^2\right) c_2^2 \\
&\kern16em-3 c_2' c_2-6 (c_2')^2 (x-1) \biggr] \,
 \frac{\del\calp}{\del\c} =0\,,
\end{split}
\eqlabel{beomx}
\end{equation}
where the prime now denotes a derivative with respect to $x$.
We demand that a physical RG flow should correspond to a background
geometry with a regular horizon.  To ensure regularity, it is
necessary to impose the following boundary conditions at the horizon:
\begin{equation}
x\to 1_- : \qquad
\biggl\{\a(x),\ \c(x),\ c_2(x)\biggr\}\longrightarrow \biggl\{\dd_1,\dd_2,\dd_3\biggr\}\,,
\eqlabel{boundh}
\end{equation}
where the $\dd_i$ are constants.

The boundary conditions at $x= 0$ are determined from the
requirement that the solution should approach the $AdS_5$ geometry
as $x\to 0_+$:
\begin{equation}
x\to 0_+ :  \qquad
\biggl\{\a(x),\ \c(x),\ c_2(x)\biggr\}\longrightarrow
\biggl\{0,0,\propto x^{-1/4}\biggr\}\,.
\eqlabel{bboun}
\end{equation}
The three supergravity parameters
$\dd_i$ uniquely determine a non-singular RG flow in the dual gauge theory.
As we review in section~\ref{sec:gtvsp}, they
are unambiguously related to the three physical parameters in
the gauge theory:
the temperature $T$, and the bosonic and fermionic masses
$m_b$ and $m_f$ of the $\caln=2^*$ hypermultiplet components.

A general analytical solution of the system \eqref{beomx} with the boundary
conditions  \eqref{boundh}, \eqref{bboun} is unknown.
However, it is possible to find an analytical solution in the regime of
high temperatures%
\footnote{Non-extremal $AdS_5$ geometry is obtained as a trivial solution for
vanishing bosonic and fermionic deformations: $\a=\chi=0$ identically.}
(see section~\ref{sec:thtpwf}).
Notice that given \eqref{pp} and \eqref{supp}, one can consistently truncate 
the supergravity system \eqref{beomx} to bosonic deformation only, \ie, by
setting $\chi=0$ identically. On the other hand, it is inconsistent (beyond
the linear approximation) to set bosonic deformation to zero, \ie, to set
$\a=0$ identically while keeping the fermionic deformation $\chi\ne 0$. 

In the rest of this section we discuss the asymptotic singularity-free
solution of \eqref{beomx} near the boundary $x\to 0_+$ and near the horizon
$x\to 1_-$, constraint by the boundary conditions \eqref{bboun} and
\eqref{boundh}, respectively. We justify referring to the $\a$ 
and $\chi$ scalars as bosonic and fermionic deformations. We also explain
the advantages of using the radial coordinate \eqref{radgauge} in numerical
integration.  Finally, we recall holographic renormalization of the thermal
PW flows.  

In what follows we find it convenient to introduce 
\begin{equation}
\begin{split}
&c_2\equiv e^A\,,\qquad A\equiv \ln\hdd_3-\frac 14 \ln (2x-x^2)+a(x)\,,\\
&\r\equiv e^\a\,.
\end{split} 
\eqlabel{defc2}
\end{equation}
The form of $A(x)$ is chosen to extract the leading asymptotic behavior
$c_2\sim x^{-1/4}$ from the metric function $c_2(x)$.  In addition, the
introduction of $\rho=e^\a$ is natural, as the bosonic scalar $\a$ enters
exponentially in the superpotential \eqref{supp}.  Imposing the boundary
conditions \eqref{boundh} and \eqref{bboun}, we see that the new warp factor
$a(x)$ and scalar $\r(x)$ satisfy the boundary conditions
\begin{equation}
\begin{split}
&x\to 0_+:\qquad a(x)\to 0\,,\qquad \r(x)\to 1\,,\\
&x\to 1_-:\qquad a(x)\to \ln\frac{\dd_3}{\hdd_3}\,,\qquad \r(x)\to e^{\dd_1}\,,
\end{split}
\eqlabel{bcax}
\end{equation}
where $\hdd_3$ is a new constant.

\subsection{Asymptotics of the thermal PW flow near the boundary}

The most general solution of \eqref{beomx} subject to the boundary conditions \eqref{bboun} and \eqref{bcax} may be expanded as a series around $x=0$:
\begin{equation}
\begin{split}
\r=\,&1+x^{1/2}\left(\r_{10}+\r_{11}\ \ln x\right)+x\left(\r_{20}+\r_{21}\ \ln x+\r_{22}\ \ln^2 x\right)\kern4em\\
&+\dots+ x^{k/2} \left(\sum_{i=1}^k \r_{ki}\ \ln^i x\right)+\cdots\,,
\end{split}
\eqlabel{bounexp1}
\end{equation}
\begin{equation}
\begin{split}
\chi=\,&\c_0 x^{1/4}\Biggl[
1+x^{1/2}\left(\c_{10}+\c_{11}\ \ln x\right)+x\left(\c_{20}+\c_{21}\ \ln x+\c_{22}\ \ln^2 x\right)\\
&+\dots+ x^{k/2} \left(\sum_{i=1}^k \c_{ki}\ \ln^i x\right)+\cdots\Biggr]\,,
\end{split}
\eqlabel{bounexp2}
\end{equation}
\begin{equation}
\begin{split}
a=\,&x^{1/2}\left(\a_{10}+\a_{11}\ \ln x\right)+x\left(\a_{20}+\a_{21}\ \ln x+\a_{22}\ \ln^2 x\right)\kern5em\\
&+\dots+ x^{k/2} \left(\sum_{i=1}^k \a_{ki}\ \ln^i x\right)+\cdots\,.
\end{split}
\eqlabel{bounexp3}
\end{equation}
This solution is characterized by five independent parameters
\begin{equation}
\biggl\{\ln\hdd_3;\ \r_{11},\r_{10};\ \c_0,\c_{10}\biggr\}\,,
\eqlabel{par}
\end{equation}
where we have included $\hdd_3$ from \eqref{defc2}.  The remaining
series coefficients are completely determined from the above parameters.

For the holographic renormalization (see subsection \ref{subsec:htftpf}) 
we will need the coefficients of the first two subleading terms in 
\eqref{bounexp1}--\eqref{bounexp3}.  Explicitly, these are given by
\begin{equation}
\begin{split}
\r_{22}=&\frac 32 \r_{11}^2\,,\\
\r_{21}=&3\r_{10}\r_{11}-8 \r_{11}^2+\frac{26}{9}\c_0^2\r_{11}\,,\\
\r_{20}=&24\r_{11}^2+\frac 32 \r_{10}^2-8\r_{10}\r_{11}+\frac{26}{9}\c_0^2\r_{10}-\frac{104}{9}\c_0^2\r_{11}+\frac 13 \c_0^4\,,\kern7.5em
\end{split}
\eqlabel{rr2}
\end{equation}
\begin{equation}
\begin{split}
\kern-17.8em\c_{11}=&\frac 13 \c_0^2\,,\kern10em
\end{split}
\eqlabel{cc1}
\end{equation}
\begin{equation}
\begin{split}
\c_{22}=&\frac 92 \r_{11}^2\,,\\
\c_{21}=&2\c_0^2\r_{11}+9\r_{10}\r_{11}+\frac{13}{36}\c_0^4-\frac{51}{2}\r_{11}^2\,,\\
\c_{20}=&\frac 18-6\c_0\r_{11}+\frac 92\r_{10}^2+\frac{243}{4}\r_{11}^2-\frac{51}{2}\r_{11}\r_{10}+2\c_0^2\r_{10}+\frac{13}{12}
\c_0^2\c_{10}-\frac{131}{120}\c_0^4\,,
\end{split}
\eqlabel{cc2}
\end{equation}
\begin{equation}
\begin{split}
a_{11}=&0\,,\\
a_{10}=&-\frac 19 \c_0^2\,,\kern27em
\end{split}
\eqlabel{aa1}
\end{equation}
\begin{equation}
\begin{split}
a_{22}=&-\frac 12 \r_{11}^2\,,\\
a_{21}=&-\r_{10}\r_{11}-\frac 12 \r_{11}^2-\frac{1}{12}\c_0^4\,,\\
a_{20}=&-\frac 12 \r_{10}\r_{11}-\frac 14 \c_0^2\c_{10}-\frac 34 \r_{11}^2+\frac{13}{648}\c_0^4-\frac 12 \r_{10}^2\,.\kern10.5em
\end{split}
\eqlabel{aa2}
\end{equation}
As indicated above, the non-extremal $AdS_5$ geometry is obtained by setting $\a\equiv 0$, $\c\equiv 0$, which in asymptotic expansions
\eqref{rr2}--\eqref{aa2} corresponds to taking $\r_{11}=\r_{10}=\c_0=0$. 
This leads to
\begin{equation}
ds_5^2 = (2\pi T)^2 \left(2x-x^2\right)^{-1/2} \left( -\left(1- x\right)^2 dt^2 + d x_1^2 +
d x_2^2 + d x_3^2 \right) + \frac{dx^2}{2x-x^2}\,,
\eqlabel{d3n}
\end{equation}
where the temperature $T$ is given by
\begin{equation}
2\pi T=\hdd_3\,.
\end{equation}
Matching with asymptotic extremal $AdS_5$ geometry with a standard Poincare
patch radial coordinate $R$ (and with our conventional choice for the radius
of curvature $L=2$)
\begin{equation}
ds_5^2 = \frac{R^2}{4} \left( - dt^2 + d x_1^2 +
d x_2^2 + d x_3^2 \right) + 4\frac{dR^2}{R^2}\,,
\eqlabel{ads5}
\end{equation}
we identify near the boundary
\begin{equation}
x\propto R^{-4}\,,\qquad x\to 0_+\,.
\eqlabel{xR}
\end{equation}
Given \eqref{xR} and the  asymptotic expansions
\eqref{bounexp1}--\eqref{bounexp3}, we identify the conformal weight two
supergravity scalar $\r(x)$ as dual to turning on mass for the bosonic
components of the $\caln=2^*$ hypermultiplet. Parameters $\r_{11}$ and
$\r_{10}$ are coefficients of its non-normalizable and normalizable 
modes, correspondingly. Similarly, the conformal weight one supergravity
scalar $\c(x)$ can be identified as dual to turning on mass for the
fermionic components of the $\caln=2^*$ hypermultiplet. Parameters $\c_{0}$
and $\c_{10}$ are coefficients of its non-normalizable and normalizable 
modes, correspondingly. We discuss the precise relation of $\r_{11}$ and
$\c_0$ to the $\caln=2^*$ gauge theory bosonic and fermionic masses in
section~\ref{sec:gtvsp}.  Here, we would simply like to emphasize the
following well-known fact \cite{ps}: given $\{\hdd, \r_{11}, \c_0\}$,
the coefficients of the normalizable modes of the supergravity scalars
$\r(x)$ and $\c(x)$, namely $\{\r_{10},\c_{10}\}$, are uniquely fixed
by requiring that the resulting supergravity RG flow is nonsingular in
the bulk. This statement is simply the supergravity dual to a gauge
theory lore: for a fixed temperature and mass parameters, the bosonic and
fermionic condensates are determined uniquely%
\footnote{Strictly speaking this is true in the absence of moduli. 
However, we do not expect moduli at finite temperature, and thus unbroken
supersymmetry, on the gauge theory side.}.
Thus, for nonsingular finite temperature Pilch-Warner flows, we must have
\begin{equation}
\r_{10}=\r_{10}(\hdd_3,\r_{11},\c_0)\,,\qquad
\c_{10}=\c_{10}(\hdd_3,\r_{11},\c_0)\,.
\eqlabel{condrel}
\end{equation}

\subsection{Asymptotics of the thermal PW flow near the horizon}
We now consider the behavior of the solution near the horizon, $x=1$.
Notice that the equations \eqref{beomx} are invariant under the transformation
\begin{equation}
(1-x)\ \Leftrightarrow\ -(1-x)\,.
\end{equation}
It is the straightforward to verify that boundary conditions \eqref{boundh},
\ie~the finiteness of $\{a(x), \r(x), \c(x)\}$ near the horizon, implies that
$\{a(x), \r(x), \c(x)\}$ are even functions of $(1-x)$ for $|1-x|\ll 1$.
It is this fact that justifies our choice of the radial coordinate $x$
in \eqref{radgauge}. Indeed, keeping fixed  $\{\hdd_3, \r_{11}, \c_0\}$,
for generic values of $\{\r_{10}, \c_{10}\}$, the functions
$\{a(x), \r(x), \c(x)\}$  would diverge at the horizon.  Thus, fixing
coefficients of the normalizable modes $\{\r_{10}, \c_{10}\}$ from 
the requirement of nonsingularity of the holographic RG flow is equivalent
to imposing Neumann boundary conditions on $\{a(x), \r(x), \c(x)\}$
at the horizon, \ie~as $x\to 1_-$
\begin{equation}
\lim_{x\to 1_-} \r'=\lim_{x\to 1_-} \c'=\lim_{x\to 1_-} a'=0\,.
\eqlabel{nbc}
\end{equation} 
The boundary condition \eqref{nbc} would determine bosonic and fermionic
condensate dependence on the hypermultiplet masses and the temperature
\eqref{condrel}.

\subsection{Holographic thermodynamics of the finite temperature PW flow}
\label{subsec:htftpf}

We use the method of holographic renormalization in order to examine the
thermodynamics of the $\caln=2^*$ RG flow.  Holographic renormalization of
this flow was explained in \cite{hr}, where it was investigated  using the
original radial coordinate $r$ of the metric ansatz \eqref{ab}. 
In particular, parameterizing
\begin{equation}
c_1(r)=e^{A(r)+B(r)}\,,\qquad c_2(r)=e^{A(r)}\,,
\eqlabel{recal1}
\end{equation}
the results of \cite{hr} indicate that the entropy density $s$, the energy
density $\cale$, and the free energy density $\calf$ are given by
\begin{equation}
\begin{split}
s=&\frac{1}{4G_5}\ \lim_{r\to r_{\rm horizon}} e^{3A}\,,\\
\cale=&\frac{1}{8\pi G_5}\ \lim_{r\to\infty}\biggl[-3e^{4A+B}A'+2e^{4A+B}\biggl\{\a_1+\a_3\ \a+\a_4\ \chi^2+\a_5\ \a^2\\
&\kern6em
+\a_6\ \a\chi^2+\a_8\ \frac{\a^2}{\ln\ee}
+\ln\ee\ \a_{10}\ \chi^4+\a_{11}\chi^4\biggr\}\biggr]\,,\\
\calf=&\cale-s T=\cale -\frac{1}{8\pi G_5}\ \lim_{r\to\infty}\biggl[e^{4A+B}B'\biggr]\,,
\end{split}
\eqlabel{efin}
\end{equation}
where 
\begin{equation}
\ee\equiv \sqrt{-g_{tt}}=e^{A+B}\,,
\eqlabel{edef}
\end{equation}
and the boundary counterterm coefficients $\a_i$ take the values
\begin{equation}
\begin{split}
\a_1&=\frac 34\,,\qquad \a_2=\frac{1}{4}\,,\qquad \a_3=0\,,\qquad 
\a_4=\frac 12\,,\qquad
\a_5=3\,,\qquad \a_6=0\,,\\
\a_7&=0\,,\qquad \a_8=-\frac 32\,,
\qquad \a_9=- \frac{1}{3}\,,\qquad \a_{10}=-\frac 23\,,\qquad  \a_{11}=\frac 16\,.
\end{split}
\eqlabel{ai1}
\end{equation}
Using the asymptotic expansions \eqref{bounexp1}--\eqref{bounexp3}, and
changing the radial coordinate in \eqref{efin} following \eqref{radgauge},
we find 
\begin{equation}
\begin{split}
\calf=-\frac{\hdd_3^4}{32\pi G_5}\biggl(
&1+\r_{11}^2 (24-96\ln\hdd_3+24\ln 2)-24 \r_{10}\r_{11}+2\c_0^2\c_{10}\\
&+\c_0^4\left(\frac 49-\frac 23\ln 2+\frac 83\ln\hdd_3\right)\biggr)\,,
\end{split}
\eqlabel{changedf}
\end{equation}
\begin{equation}
\begin{split}
\cale=\calf-\frac{1}{8\pi G_5}\hdd^4_3\,,\qquad sT=\frac {1}{8\pi G_5}\hdd^4_3\,.
\kern12em
\end{split}
\eqlabel{changede}
\end{equation}
Finally, the entropy density is given by 
\begin{equation}
s=\frac{\hdd_3^3 e^{3a_h}}{4G_5}\,,\qquad a_h=\lim_{x\to 1_-} a(x)\,,
\eqlabel{changeds}
\end{equation}
and thus we extract the temperature from \eqref{changede}
\begin{equation}
T=\frac{\hdd_3}{2\pi}\ e^{-3a_h}\,.
\eqlabel{tem}
\end{equation}

Of course, the same value of the Hawking temperature \eqref{tem} can be
extracted from the near-horizon geometry, provided one recalls that $sT$ is
a renormalization group flow invariant in the supergravity black brane
geometries without a chemical potential \cite{blu}. Indeed, from 
the relation between components of the Ricci tensor 
\begin{equation}
R_{x_1}^{\ x_1}=R_t^{\ t}\,,
\eqlabel{rel}
\end{equation}
we have a constraint%
\footnote{Eq.~\eqref{rel1} can also be directly derived from \eqref{beom}.}
\begin{equation}
c_2^4\ \left(\frac {c_1}{c_2}\right)'={\rm constant}\,.
\eqlabel{rel1}
\end{equation}
Evaluating the LHS of \eqref{rel1} near the horizon, we have 
\begin{equation}
\lim_{r\to r_{\rm horizon}}\ c_2^4\ \left(\frac {c_1}{c_2}\right)'
=8\pi G_5\ sT\,.
\eqlabel{rel2}
\end{equation}
Changing to the radial coordinate $x$ in \eqref{radgauge} and using the
asymptotic expansions \eqref{bounexp1}--\eqref{bounexp3}, the LHS of
\eqref{rel1} near the boundary takes form%
\footnote{In principle, one does not have to take a limit in \eqref{rel3}
since this expression is a constant.}
\begin{equation}
\lim_{r\to \infty}\ c_2^4\ \left(\frac {c_1}{c_2}\right)'=\lim_{x\to 0_+} \left(-\frac{dx}{dr}\right)c_2^4=\hdd_3^4\,.
\eqlabel{rel3}
\end{equation}
Comparing \eqref{rel2} with \eqref{rel3} leads to the value of $sT$ presented
in \eqref{changede}.  It is precisely the holographic RG invariance of the
combination $sT$ that guarantees the basic thermodynamic relation 
\begin{equation}
\calf=\cale-s T\,.
\eqlabel{basictermo}
\end{equation} 
Relation \eqref{basictermo} was verified explicitly for the $\caln=2^*$
thermal RG flow in \cite{bl}; the same relation holds in other thermal
gauge/gravity duals, for example in the thermal cascading RG flow \cite{aby}.

%%%%%%%%%%%%%%%%%%%%%%%%%%%%%%%%%%%%%%%%
\section{Gauge theory versus supergravity parameters}
\label{sec:gtvsp}

The relation between $N=2^*$ gauge theory and the supergravity parameters of
the (thermal) Pilch-Warner flow was established in \cite{bpp,bl}. For
completeness, we review the main points here. We begin with the gauge theory,
then move to the supersymmetric PW flow \cite{bpp}, and finally discuss the
gauge/gravity parameter identification at finite temperature. 

\subsection{$N=2^*$ gauge theory}
In the language of four-dimensional $\caln=1$ supersymmetry, the
mass deformed $\caln=4$ $SU(N)$ Yang-Mills theory ($\caln=2^*$) in
$\reals^{3,1}$ consists of a vector multiplet $V$, an adjoint chiral
superfield $\Phi$ related by $\caln=2$ supersymmetry to the gauge
field, and two additional adjoint chiral multiplets $Q$ and $\tilde{Q}$
which form an $\caln=2$ hypermultiplet.  In addition to the usual
gauge-invariant kinetic terms for these fields%
\footnote{The classical K\"{a}hler potential is normalized
according to $(2/g_{YM}^2)\tr[\bar{\Phi}\Phi+ \bar{Q}Q+\bar{\tQ}\tQ]$.},
the theory has additional interactions and a hypermultiplet mass term
given by the superpotential
\begin{equation}
W=\frac{2\sqrt{2}}{g_{YM}^2}\tr([Q,\tQ]\Phi)
+\frac{m} {g_{YM}^2}(\tr Q^2+\tr\tQ^2)\,.
\eqlabel{sp}
\end{equation}
When $m=0$, the gauge theory is superconformal with $g_{YM}$
characterizing an exactly marginal deformation. The theory has a
classical $3(N-1)$ complex dimensional moduli space, which is
protected by supersymmetry against (non)-perturbative quantum
corrections.

When $m\ne 0$, the $\caln=4$ supersymmetry is softly broken to
$\caln=2$. This mass deformation lifts the $\{Q,\ \tQ\}$ hypermultiplet
moduli directions, leaving the $(N-1)$ complex dimensional Coulomb
branch of the $\caln=2$ $SU(N)$ Yang-Mills theory, parameterized by
expectation values of the adjoint scalar
\begin{equation}
\Phi={\rm diag} (a_1,a_2,\cdots,a_N)\,,\quad \sum_i a_i=0\,,
\eqlabel{adsc}
\end{equation}
in the Cartan subalgebra of the gauge group.  For generic values
of the moduli $a_i$, the gauge symmetry is broken to that of the
Cartan subalgebra $U(1)^{N-1}$, up to the permutation of individual
$U(1)$ factors. Additionally, the superpotential \eqref{sp} induces
the RG flow of the gauge coupling.  While from the gauge theory
perspective it is straightforward to study this $\caln=2^{*}$ theory
at any point on the Coulomb branch \cite{dw}, the PW supergravity
flow \cite{pw} corresponds to a particular Coulomb branch vacuum.
More specifically, matching the probe computation in gauge theory
and the dual PW supergravity flow, it was argued in \cite{bpp} that
the appropriate Coulomb branch vacuum corresponds to a  linear
distribution of the vevs \eqref{adsc} as
\begin{equation}
a_i\in [-a_0,a_0]\,,\qquad a_0^2=\frac{m^2 g_{YM}^2 N}{\pi}\,,
\eqlabel{inter}
\end{equation}
with (continuous in the large $N$ limit) linear number density
\begin{equation}
\rho(a)=\frac{2}{m^2 g_{YM}^2}\sqrt{a_0^2-a^2}\,,\qquad
\int_{-a_0}^{a_0}da \rho(a)=N\,.
\eqlabel{rho}
\end{equation}
Unfortunately, the extension of the $N=2^*$ gauge/gravity
correspondence of \cite{pw,bpp,ejp} for vacua other than \eqref{rho}
is not known.

In \cite{bpp,ejp} the dynamics of the gauge theory on the D3 brane
probe in the PW background was studied in detail.  It was shown
in \cite{bpp} that the probe has a one complex dimensional moduli
space, with bulk induced  metric precisely equal to the metric on
the appropriate one complex dimensional submanifold of the $SU(N+1)$
$\caln=2^*$ Donagi-Witten theory Coulomb branch.  This one dimensional
submanifold is parameterized by the expectation value $u$ of the
$U(1)$ complex scalar on the Coulomb branch of the theory where
$SU(N+1)\rightarrow U(1)\times SU(N)_{PW}$. Here the $_{PW}$ subscript
denotes that the $SU(N)$ factor is in the Pilch-Warner vacuum
\eqref{rho}. Whenever $u$ coincides with any of the $a_i$ of the
PW vacuum, the moduli space metric diverges, signaling the appearance
of additional massless states.  An identical divergence is observed
\cite{bpp,ejp} for the probe D3-brane at the {\it enhancon}
singularity of the PW background.  Away from the singularity locus,
$u=a\in [-a_0,a_0]$, the gauge theory computation of the probe
moduli space metric is one-loop exact.  This is due to the suppression
of instanton corrections in the large $N$ limit \cite{bpp,b} of
$\caln=2$ gauge theories.

Consider now $\caln=2^*$ gauge theory at finite temperature $T$.
Of course, finite temperature completely breaks supersymmetry. Thus, we
can generalize the thermal $\caln=2^*$ gauge theory by allowing for different
(non-equal) masses $m_b$ and $m_f$ for the bosonic and fermionic components
of the $\caln=2^*$ hypermultiplet $\{Q,\tQ\}$ correspondingly. It is only
when $m_b=m_f=m$, and at zero temperature $T=0$, that we have $\caln=2$
supersymmetry.  Since turning on mass terms for the bosonic or fermionic
components of the hypermultiplet corresponds to deforming the $\caln=4$
supersymmetric Yang-Mills Lagrangian by relevant operators of different
classical dimension (a dimension two operator for bosonic and a dimension
three operator for fermionic mass terms), such deformations will be encoded
in different supergravity modes.  As indicated in section~\ref{sec:neg},
turning on bosonic/fermionic masses corresponds to turning on the
five-dimensional supergravity scalars $\a$/$\c$ correspondingly.

We would like to conclude this section with a simple observation.
Turning on bosonic/fermionic masses for the hypermultiplet components
sets a strong coupling scale $\Lambda \propto \max\{m_b,m_f\}$.  In this
case, we expect to find two qualitatively different phases of this gauge
theory, depending on whether $T\gg \Lambda$ or $T\ll \Lambda$.
In the former case we expect the thermodynamics to be qualitatively (and
quantitatively in the limit $T/\Lambda\to \infty$) similar to that of
the $\caln=4$ gauge theory plasma \cite{kp}. On the other hand, when
$T\sim \Lambda$, and with $m_f=0$, we actually expect an instability in the system.
Indeed\footnote{We would like to thank Ofer Aharony for clarifying this.},
turning on only the supergravity scalar $\a$, \ie, setting $m_b\ne 0$ and $m_f=0$,
corresponds to giving positive mass-squared to four out of six $\caln=4$ scalars
(these are the bosonic components of the $\caln=2$ hypermultiplet).
The remaining two $\caln=4$ scalars at the same time obtain a negative mass-squared \cite{pw} ---
they are tachyons at zero temperature. At high enough temperature the thermal corrections 
would stabilize these tachyons. However, as we lower the temperature, we expect
the re-emergence of these tachyons. As argued in \cite{gm}, dynamical
instabilities in thermal systems are reflected in thermodynamic instabilities.
Furthermore, it was argued in general (and demonstrated explicitly with a
concrete example) \cite{binst} that thermodynamic instabilities are reflected
to developing $c_s^2<0$, where $c_s$ is the speed of sound waves in the
thermal gauge theory plasma.  We discuss the supergravity realization of these
phenomena in section~\ref{sec:r}.

\subsection{Supersymmetric PW flow}

It is clear that the following conditions must hold in order to preserve
$\caln=2$ supersymmetry for the mass deformation of $\caln=4$ SYM:
\nxt the temperature must be zero: $T=0$; 
\nxt the masses for the bosonic and fermionic components of the $\caln=2$ hypermultiplet $\{Q,\tQ\}$ must be the same: $m_b=m_f=m$.\\
The former condition corresponds to a restriction
\begin{equation}
c_1(r)=c_2(r)\,,
\eqlabel{t0}
\end{equation}
ensuring Lorentz invariance of the metric.  In this case, the supergravity
RG flow cannot be parameterized by the radial coordinate $x$ introduced in
\eqref{radgauge}.  Instead, the supersymmetric flow is easiest to
parameterize in terms of the fermionic scalar $\chi\in [0,+\infty)$,
where $\chi=0$ corresponds to the asymptotic $AdS_5$ boundary, and
$\chi\to +\infty$ to the enhancon location in the bulk \cite{bpp,ejp}.  
The supersymmetric Pilch-Warner solution is then given by \cite{pw}
\begin{equation}
\begin{split}
e^A&=\frac{k \r^2}{\sinh(2\chi)}\,,\\
\r^6&=\cosh(2\chi)+\sinh^2(2\chi)\,\ln\frac{\sinh(\chi)}{\cosh(\chi)}\,,
\end{split}
\eqlabel{pwsolution}
\end{equation}
where the single integration constant $k$ is related to the hypermultiplet
mass $m$ according to \cite{bpp}
\begin{equation}
k= m L =2 m\,.
\eqlabel{kim}
\end{equation}
In order to identify the thermal RG parameters $\{\r_{11},\c_0\}$ with
the gauge theory masses, we need the boundary asymptotics of the PW flow.
This was computed in \cite{bl}.  Introducing 
\begin{equation}
\xh\equiv e^{-r/2}\,,
\eqlabel{rx}
\end{equation}
we have 
\begin{equation}
\begin{split}
\chi&= k\xh \biggl[1+ k^2\xh^2\left(\ft13+\ft{4}{3}\ln(k\xh)\right)
+k^4\xh^4\left(-\ft{7}{90}+\ft{10}{3}\ln(k\xh)+\ft{20}{9}\ln^2(k\xh)\right)
\\
&\qquad+\calo\left(k^6\xh^6\ln^3(k\xh)\right)\biggr]\,,\\
\r&= 1+ k^2\xh^2 \left(\ft13+\ft{2}{3}\ln(k\xh)\right)
+k^4\xh^4\left(\ft{1}{18}+2\ln(k\xh)+\ft23\ln^2(k\xh)\right)+\calo\left(k^6\xh^6\ln^3(k\xh)\right)\,,\\
A&= -\ln (2\xh)- \ft13 k^2\xh^2
-k^4\xh^4\left(\ft{2}{9}+\ft{10}{9}\ln(k\xh)+\ft{4}{9}\ln^2(k\xh)\right)+\calo\left(k^6\xh^6\ln^3(k\xh)\right)\,.
\end{split}
\eqlabel{uvass}
\end{equation}

\subsection{Thermal PW flow}
\label{subsec:tpwf}

We now consider the thermal PW flow with $m_b=m_f=m$.  Here we would like to
identify the parameters $\r_{11}$ and $\c_0$ for such a flow in terms of the
PW quantity $k$ (and thus, following \eqref{kim}, the hypermultiplet mass $m$)
by comparing the boundary asymptotics \eqref{bounexp1}--\eqref{bounexp3} with
\eqref{uvass}.  In order to do so, we first establish the relation between the
radial coordinate $x$ in \eqref{bounexp1}--\eqref{bounexp3} and the radial
coordinate $\hx$ in \eqref{uvass}.  Matching the asymptotic warp factors $e^A$
for the zero temperature PW flow and the thermal PW flow as $x\to 0_+$, we find
\begin{equation}
x\sim 8\hdd_3^4\ \hx^4\,,\qquad x\to 0_+\,.
\eqlabel{xxh}
\end{equation}
Next, using \eqref{xxh} and matching the leading nontrivial asymptotics
[order $x^{1/2}\ln x$ for $\r(x)$ and order $x^{1/4}$ for $\c(x)$] for
both the thermal PW solution and the supersymmetric PW flow, we find
\begin{equation}
\r_{11}=\frac{2^{1/2}\ k^2}{24\ \hdd_3^2}=\frac{\sqrt{2}}{24\pi^2}\ e^{-6a_h}\ \left(\frac mT
\right)^2\,,\qquad \c_0=\frac{2^{1/4}\ k}{2\ \hdd_3}=\frac{1}{2^{3/4}\pi}\ e^{-3a_h}\ \left(\frac mT\right)\,,
\eqlabel{match}
\end{equation} 
where we used \eqref{tem}.  We emphasize that the above matching procedure is
well-defined --- it was shown in \cite{bl} that finite temperature effects
modify the asymptotic supersymmetric PW geometry at order $\calo(x)\sim
\calo(\hx^4)$, which is subleading compared to the order at which the
matching \eqref{match} is done. 

Motivated by \eqref{match}, we now propose that in the general case,
\ie~for $m_b\ne m_f$, we may independently extract bosonic and fermionic
masses according to
\begin{equation}
\r_{11}=\frac{\sqrt{2}}{24\pi^2}\ e^{-6a_h}\ \left(\frac {m_b}{T}
\right)^2\,,\qquad \c_0=\frac{1}{2^{3/4}\pi}\ e^{-3a_h}\ \left(\frac {m_f}{T}\right)\,.
\eqlabel{match2}
\end{equation}
In section~\ref{sec:thtpwf} we recall that such an identification leads to a
consistent thermodynamics of the thermal PW flows \cite{hr,bbs} at high
temperatures. Furthermore, we later show in section~\ref{sec:r} that this
identification in fact leads to a consistent thermodynamics at any temperature.

In the rest of this subsection we derive some useful expressions for the
free energy and the speed of sound waves.  Specifically, we would like to
express the free energy and the speed of sound in terms of the coefficients
of the non-normalizable modes of the supergravity scalars $\r(x),\c(x)$. 
We also present the differential constraint on coefficients of the
normalizable modes of these supergravity scalars following from the first
law of thermodynamics.

Since we will constantly refer to the parameters $\r_{11}$ and $\c_0$
related to the bosonic and fermionic masses, we distinguish them from
derived quantities (such as the subleading coefficients) by introducing
the notation
\begin{equation}
\xi\equiv \r_{11}\,,\qquad \eta\equiv \c_{0}\,.
\eqlabel{not1}
\end{equation}
Notice that because of \eqref{match2}, $\xi$ and $\eta$ are not independent,
but are related through the ratio of bosonic and fermionic masses according
to
\begin{equation}
\eta^2=6\ \frac{m_f^2}{m_b^2}\ \xi\,.
\eqlabel{not1a}
\end{equation}

The nonsingularity of the thermal PW flow at the horizon \eqref{nbc} will determine the coefficients $\{\r_{10},\c_{10}\}$ of the normalizable modes
of the supergravity scalars $\{\r(x),\c(x)\}$ and also the value $a_h$ of
the warp factor $a(x)$ at the horizon in terms of \eqref{not1} and
\eqref{not1a}.  When $m_b\ne0$, we take
\begin{equation}
\r_{10}=\r_{10}(\xi)\,,\qquad  \c_{10}=\c_{10}(\xi)\,,\qquad a_h=a_h(\xi)\,.
\eqlabel{not2}
\end{equation}
On the other hand, for $m_b=0$, we have instead
\begin{equation}
\r_{10}=\r_{10}(\eta)\,,\qquad  \c_{10}=\c_{10}(\eta)\,,\qquad a_h=a_h(\eta)\,.
\eqlabel{not2a}
\end{equation}
Notice that a given  pair $\{\xi,\eta\}$ can be unambiguously related
to bosonic and fermionic masses, measured with respect to temperature.
Indeed, following \eqref{match2}
\begin{equation}
\left(\frac{m_b}{T}\right)^2=12\sqrt{2}\ \pi^2\ e^{6a_h}\ \xi\,,\qquad \left(\frac{m_f}{T}\right)=2^{3/4}\ \pi\ 
e^{3a_h}\ \eta\,.
\eqlabel{match3}
\end{equation}
Solving for $\hdd_3$ from \eqref{tem} and expressing $G_5$ in gauge theory variables \eqref{g5}, we find for the free energy
\begin{equation}
\begin{split}
\calf=&-\frac 18\pi^2 N^2 T^4\  e^{12a_h(\xi)}\left(1+24\xi^2\   \ln(\xi^2)\ \left(1-\frac{m_f^4}{m_b^4}\right)
+12\xi\ \frac{m_f^2}{m_b^2}\ \c_{10}(\xi)-24\xi\ \r_{10}(\xi)\right)\\
&+\calf_0\,,
\end{split}
\eqlabel{fres0}
\end{equation}
assuming $m_b\ne 0$, and
\begin{equation}
\begin{split}
\calf=&-\frac 18\pi^2 N^2 T^4\  e^{12a_h(\eta)}\left(1-\frac 23\ \eta^4  \ln(\eta^4)
+2\eta^2\ \c_{10}(\eta)\right)+\calf_0\,,
\end{split}
\eqlabel{fres00}
\end{equation}
if $\xi=0$.  In the above expressions, $\calf_0$ does not 
depend on temperature. Such a constant arises because holographic
renormalization generically%
\footnote{Notice that this is not the case when $m_b=m_f$.}
leads to $\ln T$ dependence in the free energy, which requires the
introduction of an arbitrary regularization scale $\mu$
\begin{equation}
\ln T \ \to \ln \frac T\mu\,.
\eqlabel{muintr}
\end{equation}
Changing this scale modifies $\calf_0$, but otherwise has no effect on the
thermodynamics.  This phenomena has been encountered previously; it was
discussed in perturbative gauge theory thermodynamics in \cite{pertg}, and in
the context of gauge/string theory correspondence in \cite{npertg}%
\footnote{It appears even in the high temperature thermodynamics of the
thermal PW flow \cite{bbs} (see also section~\ref{sec:thtpwf}).}.
Notice that in \eqref{fres0} we wrote $\ln(\xi^2)$ instead of $2\ln (\xi)$.
The reason is that the former expression allows us to study thermodynamics for
$\xi<0$, which according to \eqref{match2} we interpret as $m_b^2<0$.
While introducing tachyonic masses for gauge theory scalars at zero
temperature leads to instability, this is not the case at finite temperature.
Here, the effective mass squared receives thermal corrections of order $T^2$,
which might cure the zero temperature instability. 

Besides the free energy, we can also evaluate the entropy density and
the energy density
\begin{equation}
s= \frac 12\pi ^2 N^2 T^3\ e^{12 a_h}\,,\qquad \cale=\calf+sT \,.
\eqlabel{seres}
\end{equation}
In addition, from the defining equations \eqref{match3}, we can evaluate
the quantities $d\xi/dT$ and $d\eta/dT$:
\begin{equation}
\begin{split}
&\frac{d\xi}{dT}=-\frac{m_b^2}{6\sqrt{2}\pi^2 T^3}\ e^{-6 a_h(\xi)}\ \left(6\xi\ \frac{d a_h(\xi)}{d\xi}+1\right)^{-1}\,,
\qquad m_b\ne 0\,;\\
&\frac{d\eta}{dT}=-\frac{m_f}{2^{3/4}\pi T^2}\ e^{-3 a_h(\eta)}\ \left(3\eta\ \frac{d a_h(\eta)}{d\eta}+1\right)^{-1}\,,
\qquad m_b= 0\,.\\
\end{split}
\eqlabel{defder}
\end{equation}
Using \eqref{fres0} or \eqref{fres00}, \eqref{seres} and \eqref{defder} the first law of thermodynamics,
\begin{equation}
d\calf=-s\ dT\,,
\eqlabel{deff}
\end{equation}
reduces to a differential constraint on the quantities%
\footnote{Notice that even though, for $\xi=0$, the free 
energy density \eqref{fres00} does not depend on 
$\r_{10}$, one generically expects that $\r_{10}=\r_{10}(\eta)$.}
\begin{equation}
\begin{split}
&\{\xi,\c_{10}(\xi),\r_{10}(\xi),a_h(\xi)\}\qquad\hbox{for }m_b\ne0\,;\\
&\{\eta,\c_{10}(\eta),\r_{10}(\eta),a_h(\eta)\}\qquad\hbox{for }m_b=0\,.
\end{split}
\end{equation}
In particular, the first law in differential form reads
\begin{equation}
\begin{split}
0=&4\xi\left(1-\frac{m_f^4}{m_b^4}\right)+2\left(\r_{10}(\xi)-\xi\frac{d\r_{10}(\xi)}{d\xi}\right)
-\frac{m_f^2}{m_b^2}\left(\c_{10}(\xi)-\xi\frac{d\c_{10}(\xi)}{d\xi}\right)+\frac{da_h(\xi)}{d\xi}\,,
\end{split}
\eqlabel{firstlaw}
\end{equation}
for $m_b\ne 0$, and 
\begin{equation}
0=\frac 43 \eta^3+2\eta\ \c_{10}(\eta)-\eta^2\ \frac{d\c_{10}(\eta)}{d\eta}-6\frac{da_h(\eta)}{d\eta}\,,
\eqlabel{firstlawf}
\end{equation}
for $m_b=0$.  Using either \eqref{firstlaw} or \eqref{firstlawf}, we
can evaluate the speed of sound entirely as a function of $\xi$ or $\eta$.
Straightforward algebraic manipulations result in 
\begin{equation}
c_s^2=\frac{\del P}{\del \cale}=-\frac{\del \calf}{\del \cale}=\frac{1+\Sigma}{3-\Sigma}\,,
\eqlabel{vs2}
\end{equation}
where
\begin{equation}
\begin{split}
\Sigma\equiv \Sigma(\xi)=&24\xi^2\ \left(\frac{m_f^4}{m_b^4}-1\right)+12\xi\left(\xi\frac{d\r_{10}(\xi)}{d\xi}-\r_{10}(\xi)\right)
\\
&+6\xi\ \frac{m_f^2}{m_b^2}\ \left(\c_{10}(\xi)-\xi\frac{d\chi_{10}(\xi)}{d\xi}\right)\,,\qquad m_b\ne 0\,;\\
\Sigma\equiv \Sigma(\eta)=&\frac 23 \eta^4+\eta^2\ \chi_{10}(\eta)-\frac 12\ \eta^3\ \frac{d\chi_{10}(\eta)}{d\eta}
\,,\qquad m_b=0\,,
\end{split}
\end{equation}
and where $P=-\calf$ is the pressure. 

In general, one may explore the PW thermodynamics in the entire phase plane
of $m_b$-$m_f$.  However, here we focus on two special cases:
\nxt when $m_f=0$ (we call this the {\it bosonic} case)%
\footnote{Recall that this corresponds to a consistent truncation of the
supergravity}
we have
\begin{equation}
\calf_{bosonic}=-\frac{\pi^2N^2 T^4}{8}\ e^{12a_h(\xi)}\  \biggl(1+24\xi^2\ln(\xi^2)-24\xi\ \r_{10}(\xi)\biggr)+\calf_0\,,
\eqlabel{mff}
\end{equation}
\begin{equation}
\Sigma_{bosonic}(\xi)=-24 \xi^2-12\xi\ \r_{10}(\xi) +12 \xi^2\ \frac{d\r_{10}(\xi)}{d\xi}\,,
\eqlabel{speedb}
\end{equation} 
and the first law of thermodynamics takes form
\begin{equation}
0=4\xi+2\left(\r_{10}(\xi)-\xi\frac{d\r_{10}(\xi)}{d\xi}\right)
+\frac{da_h(\xi)}{d\xi}\,;
\eqlabel{1stb}
\end{equation}
\nxt when $m_f=m_b\equiv m$ (we call this the {\it supersymmetric} case), or
correspondingly from \eqref{not1a}
\begin{equation}
\eta^2=6\xi\,,
\eqlabel{susy}
\end{equation}
we have 
\begin{equation}
\begin{split}
\calf_{susy}=&-\frac{\pi^2N^2 T^4}{8}\ e^{12a_h(\xi)}\  \biggl(1-24\xi\ \r_{10}(\xi)+12\xi\ \c_{10}(\xi)\biggr)+\calf_0\,,
\end{split}
\eqlabel{mfs}
\end{equation}
\begin{equation}
\Sigma_{susy}(\xi)=12\xi\left(\xi\frac{d\r_{10}(\xi)}{d\xi}-\r_{10}(\xi)\right)
+6\xi\ \left(\c_{10}(\xi)-\xi\frac{d\chi_{10}(\xi)}{d\xi}\right)\,,
\eqlabel{speeds}
\end{equation} 
and the first law of thermodynamics takes form
\begin{equation}
0=2\left(\r_{10}(\xi)-\xi\frac{d\r_{10}(\xi)}{d\xi}\right)
-\left(\c_{10}(\xi)-\xi\frac{d\c_{10}(\xi)}{d\xi}\right)
+\frac{da_h(\xi)}{d\xi}\,.
\eqlabel{1sts}
\end{equation}
Notice that in the supersymmetric case there is no $\ln(\xi^2)$ dependence
in the holographic free energy \eqref{fres0}.  As a result one can
unambiguously determine $\calf_0$ directly from \eqref{efin}%
\footnote{The latter holographic prescription was formulated in such a way
that the holographic energy of the supersymmetric PW flow vanishes \cite{hr}.}.
We find
\begin{equation}
\calf_0\equiv\calf_0^{susy}=-\frac{\pi^2N^2 T^4}{8}\ e^{12a_h(\xi)}\ 40\xi^2=-\frac{5N^2}{288\pi^2}\ m^4\,.
\eqlabel{fosusy}
\end{equation}
%

%%%%%%%%%%%%%%%%%%%%%%%%%%%%%%%%%%%%%%%%
\section{Thermodynamics of the high temperature Pilch-Warner flow}
\label{sec:thtpwf}

We now recall that the differential equations \eqref{beomx} describing the
finite temperature PW renormalization group flow admit a perturbative
analytical solution at high temperature \cite{bl,hr,bbs}. In this regime,
the appropriate expansion parameters are
\begin{equation}
\dd_1\propto \left(\frac{m_b}{T}\right)^2\ll1\,,
\qquad \dd_2\propto \frac{m_f}{T}\ll 1\,.
\eqlabel{larget}
\end{equation}
In this case, the metric function $A(x)$ is given to first nontrivial order
in $\dd_1$, $\dd_2$ by
\begin{equation}
\begin{split}
A(x)=&\ln\dd_3-\frac 14\ \ln \left(2x-x^2\right)+\dd_1^2\ A_1(x)+\dd_2^2\ A_2(x)\,,\\
\a(x)=&\dd_1\ \a_1(x)\,,\\
\c(x)=&\dd_2\ \c_2(x) \,,
\end{split}
\eqlabel{hightsol}
\end{equation}
where
\begin{equation}
\a_1=\left(2x -x^2\right)^{1/2}\ _2 F_1\left(\ft 12,\ft 12; 1; (1-x)^2\right)\,,
\eqlabel{orko1}
\end{equation}
\begin{equation}
\c_2=(2x-x^2)^{3/4}\ _2F_1\left(\ft 34, \ft 34; 1; (1-x)^2\right)\,,
\eqlabel{orko2}
\end{equation}
\begin{equation}
\begin{split}
A_1=&4\int_x^1\ \frac{(z-1)dz}{(2z-z^2)^2}
\left(\ga_1-\int_z^1dy\left(\frac{\del\a_1}{\del y}
\right)^2\frac{(2y-y^2)^2}{y-1}\right)\,,\\
A_2=&\frac 43\int_x^1\ \frac{(z-1)dz}{(2z-z^2)^2}
\left(\ga_2-\int_z^1dy\left(\frac{\del\c_2}
{\del y}\right)^2\frac{(2y-y^2)^2}{y-1}\right)\,.\\
\end{split}
\eqlabel{expsol}
\end{equation}
The constants $\ga_i$ are fine-tuned to satisfy the boundary
conditions, and are given by
\begin{equation}
\ga_1=\frac{8-\pi^2}{2\pi^2}\,,\qquad \ga_2=\frac{8-3\pi}{8\pi}\,.
\eqlabel{gai}
\end{equation}

Comparing \eqref{orko1}, \eqref{orko2}  and \eqref{bounexp1}, \eqref{bounexp2}
we find the relations
\begin{equation}
\begin{split}
&\r_{11}=-\frac{\sqrt{2}}{\pi}\ \dd_1\,,\qquad \r_{10}=\frac{3\sqrt{2}\ \ln 2 }{\pi}\ \dd_1\,,\\
&\c_0=\frac{2^{1/4}\sqrt{\pi}}{\Gamma\left(\frac 34\right)^2}\ \dd_2\,,\qquad \c_{10}=
-\frac{2^{1/2}\Gamma\left(\frac 34\right)^4}{\pi^2} \,.
\end{split}
\eqlabel{rel6}
\end{equation}
Also notice that 
\begin{equation}
\begin{split}
2\pi T=&\dd_3\left(1+\dd_1^2\left(2-\frac{d^2A_1}{dx^2}\bigg|_{x=1}\right)+\dd_2^2\left(\frac 12
-\frac{d^2A_2}{dx^2}\bigg|_{x=1}\right)\right)\\
=&\dd_3\left(1+\frac{16}{\pi^2}\ \dd_1^2+\frac{4}{3\pi}\ \dd_2^2\right) \,.
\end{split}
\eqlabel{rel7}
\end{equation}
The parameters $\{\dd_1,\dd_2\}$ can be related to the
$\left\{m_b/T,m_f/T\right\}$ parameters of the dual gauge theory via
\begin{equation}
\begin{split}
\dd_1=&-\frac{1}{24\pi}\ \left(\frac{m_b}{T}\right)^2\,,\\
\dd_2=&\frac{\left[\Gamma\left(\ft 34\right)\right]^2}{2\pi^{3/2}}
\ \frac{m_f}{T}\,.
\end{split}
\eqlabel{ddphys}
\end{equation}

To leading order in mass deformation, the free energy \eqref{fres0} takes
form \cite{hr,bbs}
\begin{equation}
\begin{split}
\calf=-\frac 18 \pi^2 N^2 T^4\left[ 1-\frac{192}{\pi^2}\ \ln(\pi T)\
\delta_1^2-\frac{8}{\pi}\ \delta_2^2\right]\,.
\end{split}
\eqlabel{pert}
\end{equation}
One can further evaluate the entropy density of the non-extremal PW geometry,
\begin{equation}
s=\frac 12 \pi^2 N^2 T^3\left(1-\frac{48}{\pi^2}\ \delta_1^2-\frac{4}{\pi}\ \delta_2^2\right)\,,
\end{equation}
and verify that the first law of thermodynamics \eqref{deff} is satisfied.

%%%%%%%%%%%%%%%%%%%%%%%%%%%%%%%%%%%%%%%%
\section{Numerical procedure and results}
\label{sec:np}

We are mainly interested in the thermodynamics of the PW flow away
from the high temperature limit.  In this case, since analytic solutions
to the equations of motion \eqref{beomx} are unavailable, we resort to
numerical integration.  In this section we describe the numerical procedure
and some of the consistency checks we have performed on the numerical
results.  Extracting the actual thermodynamics from these solutions will be
taken up in the following section.

The numerical procedure is conceptually the same for both the bosonic and 
supersymmetric mass deformations of the non-extremal $AdS_5$ geometry.
In practice, however, it is much easier to implement for the
bosonic deformation
\begin{equation}
m_f=0\,,\qquad m_b\ne 0\,,
\eqlabel{mbos}   
\end{equation}
and so we will describe the details of this first.  We then outline the
modifications required for handling the supersymmetric mass deformation
\begin{equation}
m_f=m_b\equiv m\,,
\eqlabel{msus}
\end{equation}
and comment on thermal flows corresponding to generic mass deformations. 
Finally, we discuss the validity of the supergravity approximation. 

\subsection{Bosonic deformation}
\label{subsec:bd}

Simplicity of treating the bosonic mass deformation stems from the
fact that thermal PW flows allow for a consistent truncation with the
supergravity scalar $\chi(x)$ identically set to zero.  The relevant
equations can thus be obtained from 
\eqref{beomx} with
\begin{equation}
\chi(x)\equiv 0\,,\qquad \a\equiv \ln\r\,,\qquad \ln c_2(x)\equiv A(x)\equiv \ln\hdd_3-\frac 14\ln(2x-x^2)+a(x)\,.
\eqlabel{bose0}
\end{equation}
The boundary conditions (which guarantee that the flow is singularity-free)  
are given by \eqref{bcax} and \eqref{nbc}
\begin{equation}
\begin{split}
&x\to 0_+\qquad a(x)\to 0\,,\qquad \r(x)\to 1\,; \\
&x\to 1_-\qquad a'(x)\to 0\,,\qquad \r'(x)\to 0\,.
\end{split}
\eqlabel{bose1}
\end{equation}
Asymptotic expansions for $\{\r(x),a(x)\}$ as $x\to 0_+$ are given by \eqref{bounexp1}, \eqref{bounexp3}. 
The first couple terms are presented in \eqref{rr2}-\eqref{aa2}, where one has to set $\c_0=0$
corresponding to $\chi(x)\equiv 0$, see \eqref{bounexp2}. 
Notice that numerical integration of $\r(x)$ and $a(x)$ from $x=0_+$ is uniquely determined 
specifying the non-normalizable  $\r_{11}$ and normalizable $\r_{10}$ mode coefficients.
However, a generic choice of a pair $\{\r_{11},\r_{10}\}$ would produce a solution that does not 
satisfy the horizon boundary condition (the $x\to 1_-$ boundary condition  in \eqref{bose1}).
Solving Sturm-Liouville problem with boundary conditions \eqref{bose1} would determine
\begin{equation}
\r_{10}\equiv\r_{10}(\r_{11})\qquad \Longleftrightarrow \qquad \r_{10}\equiv \r_{10}(\xi) \,,
\eqlabel{bose2} 
\end{equation}  
where on the RHS we used conventions of subsection \ref{subsec:tpwf}.
For a flow satisfying \eqref{bose2} we can extract
\begin{equation}
a_h(\xi)=\lim_{x\to 1_-} a(x) \,.
\eqlabel{bose3}
\end{equation}
To summarize, each  non-singular bosonic deformation flow would generate a triplet of numbers
\begin{equation}
\{\xi, \r_{10}(\xi), a_h(\xi)\}\,,
\eqlabel{bose4}
\end{equation}
which can be used to evaluate  the free energy $\calf_{bosonic}$ \eqref{mff}, the speed of sound $c_{s,bosonic}^2$ \eqref{vs2} 
and \eqref{speedb}, and verify (numerically) the first law of thermodynamics \eqref{1stb}. 

We use the following algorithm to generate triplets \eqref{bose4}.   
First, we choose a mass deformation parameter $\r_{11}=\xi$ which we keep 
fixed during the iteration procedure. The iteration starts by choosing a trial value of $\r_{10}$. 
Given $\{\r_{11}, \r_{10}\}$ we integrate numerically%
\footnote{We used $Mathematica$ for numerical integration as well as our
own $C$ code based on fifth-order Runge-Kutta.  Both procedures (up to
controllable numerical errors) produced equivalent results.}
\eqref{beomx} (with \eqref{bose0}) from 
$x_{initial}\ll 1$ to $x_{final}$ (such that $(1-x_{final})\ll 1$) using the power series \eqref{bounexp1},
\eqref{bounexp2} to specify\footnote{In practice we truncated the power series to $k=3$ terms (inclusive), 
and used $x_{initial}=10^{-9}$, $x_{final}=1-10^{-6}$.} 
\begin{equation}
\rho\left(x_{initial}\right)\,, \qquad \r'\left(x_{initial}\right)\,, \qquad a\left(x_{initial}\right)\,,\qquad a'\left(x_{initial}\right)\,.
\end{equation}
For a generic value of $\r_{10}$, 
\begin{equation}
\lim_{x\to 1_{-}}\rho\left(x\right)\to \pm \infty\,,\qquad \lim_{x\to 1_{-}}\a\left(x\right)\to -\infty\,.
\end{equation} 
We find that for a given value $\r_{11}=\xi$ there is  (generically) a unique $\r_{10}=\r_{10}(\xi)$, such that $\r(x)$ and $a(x)$ 
are finite as we go to the horizon, $x\to 1_{-}$, \ie, 
\begin{equation}
\lim_{x\to 1_-}{\frac{d\r}{dx}(x; \r_{11}=\xi,\hr_{10}=\r_{10}(\xi))}=0\,,
\eqlabel{hit}
\end{equation}
while 
\begin{equation}
\lim_{x\to 1_-}{\frac{d\r}{dx}(x; \r_{11}=\xi,\r_{10}\ne \r_{10}(\xi))}=\pm \infty\,.
\eqlabel{nothit}
\end{equation}
Now, the iterative algorithm is clear: we use Newton's shooting method of determining $\r_{10}(\xi)$ for a fixed $\r_{11}=\xi$. 

\begin{figure}[t]
  \hspace*{-20pt}
\psfrag{rho11}{\raisebox{1ex}{\footnotesize\hspace{-0.8cm}$\qquad \xi$}}
\psfrag{ah}{\raisebox{0.0ex}{$a_h(\xi)$}}
\psfrag{rc}{\raisebox{0.0ex}{$\r_{crit}$}}
\psfrag{rho10}{\raisebox{0.5ex}{$\r_{10}(\xi)$}}
  \includegraphics[width=3.4in]{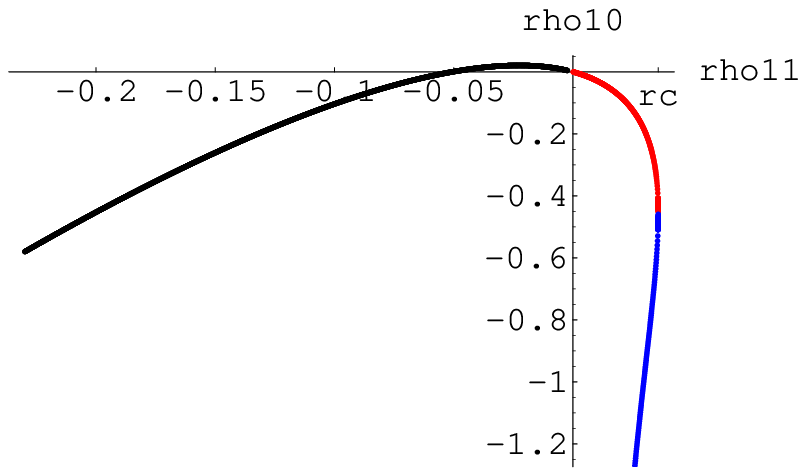}
%  \hspace*{-10pt}
  \includegraphics[width=3.4in]{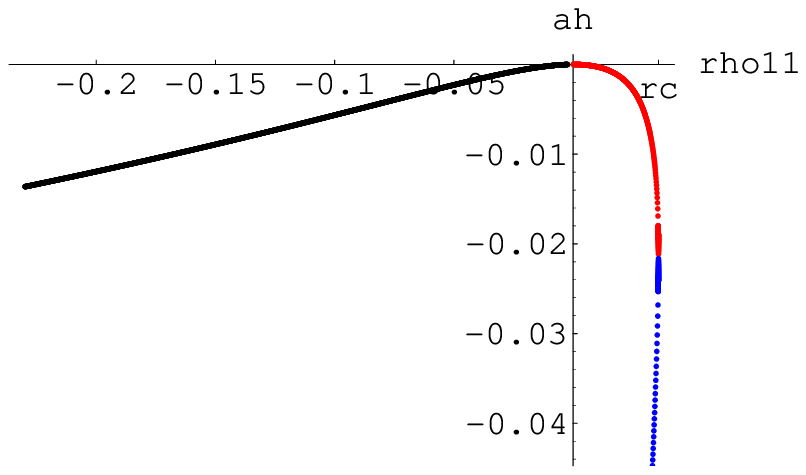}%
%  \hspace*{-20pt}
%  \vspace*{-10pt}
  \caption{
Altogether we obtained 3227 triplets $\{\xi,\r_{10}(\xi),a_h(\xi)\}$.  
The left plot represents $\r_{10}(\xi)$, and the right plot represents $a_h(\xi)$. The critical value of $\xi$ is at
$\r_{crit}\approx 0.03582(2)$. Red segments of the plots correspond to $\xi\in[0,\r_{crit})$ such that $\r_{10}(0)=a_h(0)=0$.
Blue segments of the plots do not pass through the origin and  correspond to  $0<\xi <\r_{crit}$. Black segments 
of the plots correspond to $\xi<0$. 
}
\label{figb1}
\end{figure}

\begin{figure}[t]
  \hspace*{-20pt}
\psfrag{rho11}{\raisebox{1ex}{\footnotesize\hspace{-0.8cm}$\qquad \xi$}}
\psfrag{dah}{\raisebox{0.0ex}{$a_h'(\xi)$}}
\psfrag{rc}{\raisebox{0.0ex}{$\r_{crit}$}}
\psfrag{drho10}{\raisebox{0.5ex}{$\r_{10}'(\xi)$}}
\includegraphics[width=3.4in]{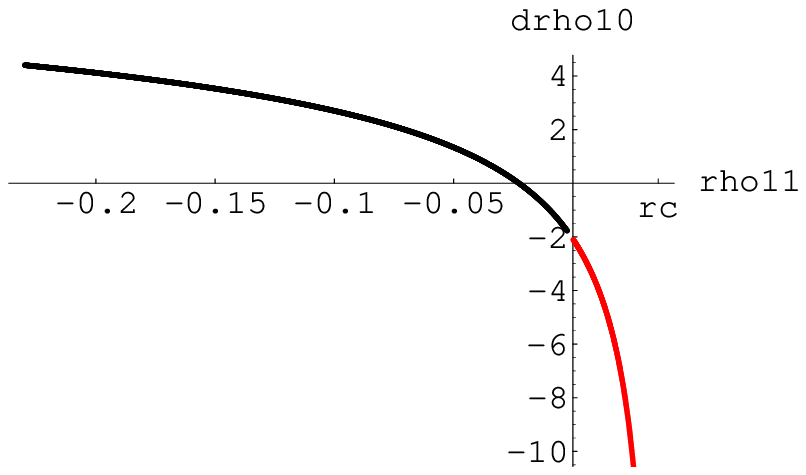}
%  \hspace*{-10pt}
  \includegraphics[width=3.4in]{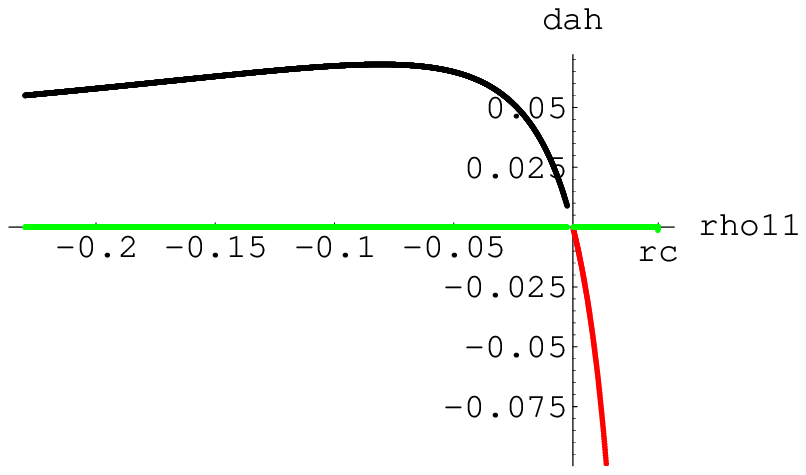}%
%  \hspace*{-20pt}
%  \vspace*{-10pt}
  \caption{
The left plot represents $\r_{10}'(\xi)$ for both the red and the black segments of the left plot in Fig.~\ref{figb1}.
The right plot represents $a_h'(\xi)$ for both the red and the black segments of the right plot in Fig.~\ref{figb1}.
The green points on the right plot represent numerical verification of the first law of the thermodynamics, 
where $a_h'(\xi)$ must be canceled with an appropriate combination of $\{\xi,\r_{10}(\xi),\r'(\xi)\}$, see \eqref{1stb}.  
}
\label{figb2}
\end{figure}

\begin{figure}[t]
  \hspace*{-20pt}
\psfrag{rho11}{\raisebox{1ex}{\footnotesize\hspace{-0.8cm}$\qquad \xi$}}
\psfrag{dah}{\raisebox{0.0ex}{$a_h'(\xi)$}}
\psfrag{rc}{\raisebox{0.0ex}{$\r_{crit}$}}
\psfrag{drho10}{\raisebox{0.5ex}{$\r_{10}'(\xi)$}}
\includegraphics[width=3.4in]{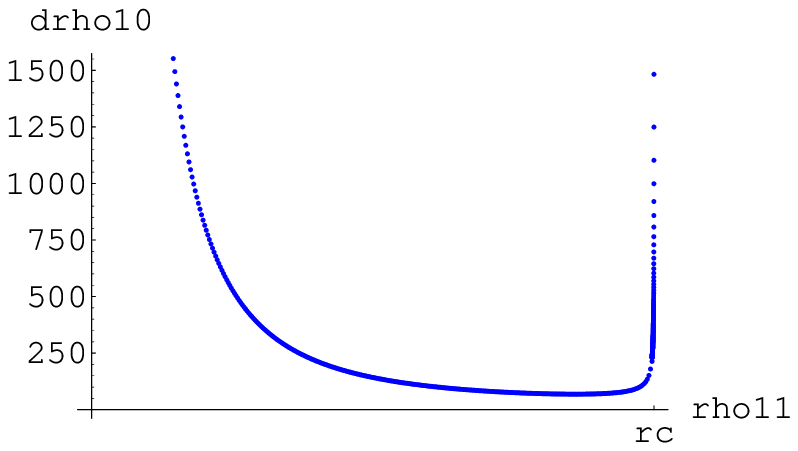}
%  \hspace*{-10pt}
  \includegraphics[width=3.4in]{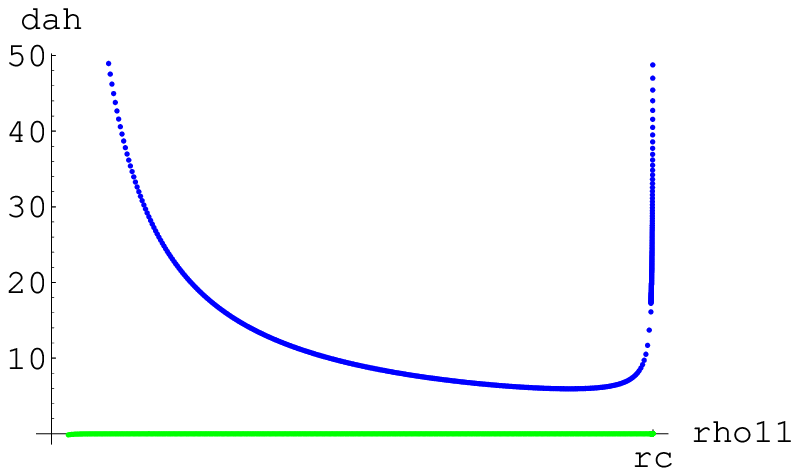}%
%  \hspace*{-20pt}
%  \vspace*{-10pt}
  \caption{
The left plot represents $\r_{10}'(\xi)$ for  the blue segment of the left plot in Fig.~\ref{figb1}.
The right plot represents $a_h'(\xi)$ for the blue segment of the right plot in Fig.~\ref{figb1}.
The green points on the right plot represent numerical verification of the first law of the thermodynamics, 
where $a_h'(\xi)$ must be canceled with an appropriate combination of $\{\xi,\r_{10}(\xi),\r'(\xi)\}$, see \eqref{1stb}. 
}
\label{figb3}
\end{figure}

Results of the numerical analysis are presented in Figs.~\ref{figb1}-\ref{figb3}. 
From Fig.~\ref{figb1} notice that for $\r_{11}> \r_{crit}\approx 0.03582$ there is no solution to Sturm-Liouville problem with boundary 
condition \eqref{bose1}. Using \eqref{match3} we find that corresponding to $\r_{crit}$,
\begin{equation}
\frac{m_b}{T}\bigg|_{crit}\approx2.29(9)\,.
\eqlabel{mbcrit}
\end{equation} 
We find that for each value of $\xi \in (0,\r_{crit})$ there are two sets $\{\r_{10}(\xi),a_h(\xi)\}$ which are represented 
by red/blue segments of the plots in  Fig.~\ref{figb1}. For each $\xi<0$ we find a single set  $\{\r_{10}(\xi),a_h(\xi)\}$ 
represented by black segments of the plots in Fig.~\ref{figb1}.
In section \ref{sec:r} we show that red/blue values of the bosonic condensate $\r_{10}(\xi)$ correspond to
different phases of the $\caln=2^*$ plasma.  We also show that there is $\xi=\rho^*<\r_{crit}$ such that 
for all $\xi>\r^*$,   $\caln=2^*$ plasma becomes unstable both thermodynamically and dynamically, in agreement 
with general  arguments of \cite{binst}.

Figs.~\ref{figb2}-\ref{figb3} represent $\r_{10}'(\xi)$, $a_h'(\xi)$, and a numerical verification of the first law of thermodynamics 
\eqref{1stb}. Since our numerical data is rather densely spaced\footnote{Mostly we have $\Delta\xi=10^{-4}$.} in $\xi$, we use 
\begin{equation}
f'(\xi)=\frac{df(\xi)}{d\xi}=\frac{f(\xi+\Delta\xi)-f(\xi-\Delta\xi)}{2\Delta\xi}\,,
\eqlabel{derf}
\end{equation}
as a numerical approximation for a derivative of a function $f(\xi)$.
Notice that  
\begin{equation}
\lim_{\xi\to \r_{crit}-0}\frac{d\r_{10}(\xi)}{d\xi}=\mp \infty\,,\qquad \lim_{\xi\to \r_{crit}-0}\frac{da_h(\xi)}{d\xi}=\mp \infty\,,
\eqlabel{rdercr}
\end{equation}
with minus/plus signs for red/blue segments of the curves in Fig.~\ref{figb2}/Fig.~\ref{figb3} correspondingly.
Also, from Fig.~\ref{figb3}
\begin{equation}
\lim_{\xi\to +0}\frac{d\r_{10}(\xi)}{d\xi}=+\infty\,,\qquad \lim_{\xi\to +0}\frac{da_h(\xi)}{d\xi}=+\infty\,.
\eqlabel{rder0}
\end{equation}
We find that 
\begin{equation}
\r_{10}'(\xi)\bigg|_{\xi=+2\times 10^{-4}}\approx -2.10(6)\,,
\eqlabel{der}
\end{equation}
which is rather close to a predicted value from the high temperature analysis of section \ref{sec:thtpwf} (see \eqref{rel6})
\begin{equation}
\r_{10}'(\xi)\bigg|_{\xi=0}^{high\ temperature}=-3\ln 2\approx -2.07944\,.
\eqlabel{predr10}
\end{equation}
A better agreement is achieved if we fit the first 200 points of the red segment of  $\r_{10}(\xi)$ 
on the left plot in Fig.~\ref{figb1} with a 10th order polynomial in $\xi$. We find in this case 
\begin{equation}
\r_{10}'(\xi)\bigg|_{\xi=0}^{fit}\approx -2.07972\,.
\eqlabel{fitr10}
\end{equation}
The green points on the right plots in Figs.~\ref{figb2}-\ref{figb3} demonstrate the cancellation of $a'_h(\xi)$ with a combination
$$
4\xi+2\left(\r_{10}(\xi)-\xi \frac{d\r_{10}(\xi)}{d\xi}\right)\,,
$$
as required by the first law of thermodynamics, see \eqref{1stb}. We find that such a cancellation is achieved with an accuracy of better 
than $10^{-4}$. Clearly, this provides a rather impressive check on both the validity of the holographic renormalization 
of the thermal PW flows explained in \cite{hr}, and the numerical procedures developed in this paper.

\subsection{Supersymmetric deformation}
Albeit more complicated, the Sturm-Liouville problem for the supersymmetric mass deformation is conceptually 
similar to the one for the bosonic mass deformation studied in previous subsection. Here we have to use all 
equations in \eqref{beomx}. 
As for the bosonic deformation, we rewrite these equations using $\{\r(x),a(x)\}$ such that 
\begin{equation}
\a\equiv \ln\r\,,\qquad \ln c_2(x)\equiv A(x)\equiv \ln\hdd_3-\frac 14\ln(2x-x^2)+a(x)\,.
\eqlabel{susy0}
\end{equation}
The boundary conditions (which guarantee that the flow is singularity-free)  
are given by \eqref{bcax} and \eqref{nbc}
\begin{equation}
\begin{split}
&x\to 0_+\qquad a(x)\to 0\,,\qquad \r(x)\to 1\,,\qquad \c(x)\to 0\,; \\
&x\to 1_-\qquad a'(x)\to 0\,,\qquad \r'(x)\to 0\,,\qquad \c'(x)\to 0\,.
\end{split}
\eqlabel{susy1}
\end{equation}
For the bosonic mass deformation we have a one-dimensional Sturm-Liouville problem: given $\r_{11}$, 
the boundary condition at the horizon determines $\r_{10}$; the pair $\{\r_{11}\equiv \xi,\r_{10}(\xi)\}$ 
further determines the value of the $a(x)$ at the horizon, $a_h(\xi)$.  For the supersymmetric 
mass deformation the Sturm-Liouville problem is two-dimensional: given $\r_{11}\equiv \xi$, 
the boundary condition \eqref{susy0} would determine
\begin{equation}
\r_{10}\equiv \r_{10}(\xi)\,,\qquad \c_{10}\equiv \c_{10}(\xi)\,.
\eqlabel{susy2}
\end{equation}
For a flow satisfying \eqref{susy2} we can extract 
\begin{equation}
a_h(\xi)=\lim_{x\to 1_-} a(x)\,.
\eqlabel{susy3}
\end{equation} 
Thus, each non-singular supersymmetric mass deformation flow would generate a set of four 
 numbers 
\begin{equation}
\{\xi,\r_{10}(\xi),\chi_{10}(\xi),a_h(\xi)\}\,,
\eqlabel{susy4}
\end{equation}
which can be used to evaluate the free energy $\calf_{susy}$ \eqref{mfs}, the speed of sound 
$c_{s,susy}^2$ \eqref{vs2} and \eqref{speeds}, and verify (numerically) the first law of thermodynamics 
\eqref{1sts}.

\begin{figure}[t]
  \hspace*{-20pt}
\psfrag{rho11}{\raisebox{1ex}{\footnotesize\hspace{-0.8cm}$\qquad \xi$}}
\psfrag{ah}{\raisebox{0.0ex}{$a_h(\xi)$}}
\psfrag{chi10}{\raisebox{0.5ex}{$\c_{10}(\xi)$}}
\psfrag{rho10}{\raisebox{0.5ex}{$\r_{10}(\xi)$}}
  \includegraphics[width=3.4in]{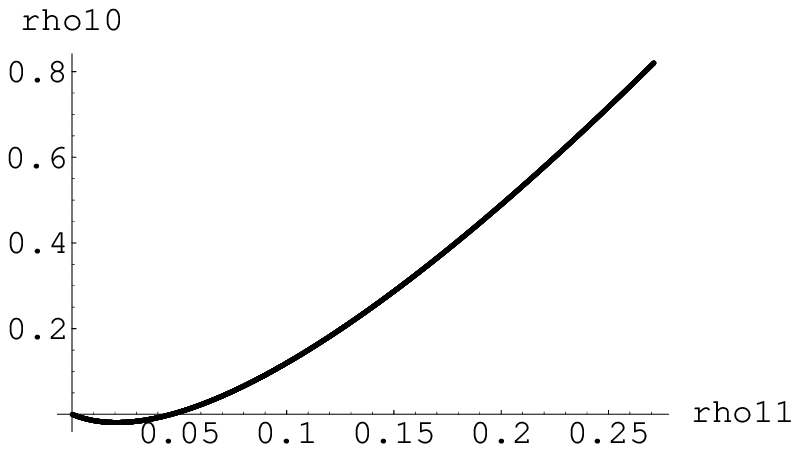}
%  \hspace*{-10pt}
  \includegraphics[width=3.4in]{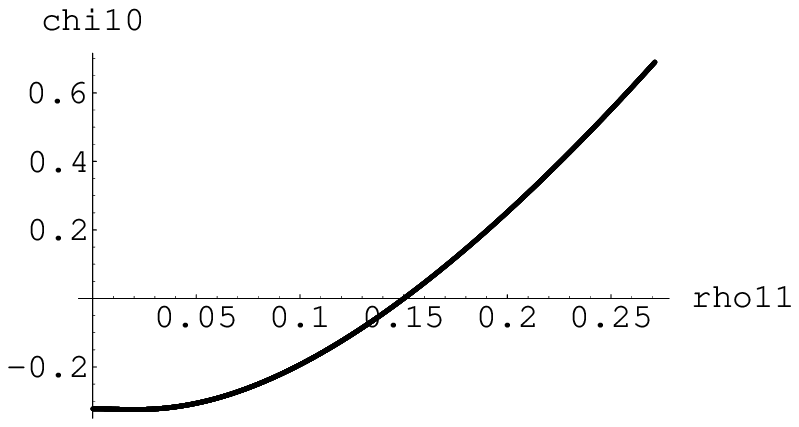}%
%  \hspace*{-20pt}
%  \vspace*{-10pt}
  \caption{
Altogether we obtained 2712 sets $\{\xi,\r_{10}(\xi),\chi(\xi),a_h(\xi)\}$.  
The left plot represents $\r_{10}(\xi)$, and the right plot represents $\chi_{10}(\xi)$. 
}
\label{figs1}
\end{figure}

\begin{figure}[t]
  \hspace*{-20pt}
\psfrag{rho11}{\raisebox{1ex}{\footnotesize\hspace{-0.8cm}$\qquad \xi$}}
\psfrag{ah}{\raisebox{0.0ex}{$a_h(\xi)$}}
\psfrag{chi10}{\raisebox{0.5ex}{$\c_{10}(\xi)$}}
\psfrag{rho10}{\raisebox{0.5ex}{$\r_{10}(\xi)$}}
  \includegraphics[width=3.4in]{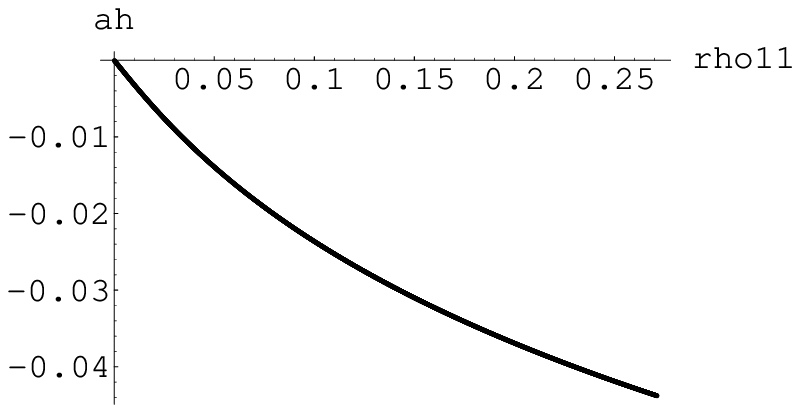}
%  \hspace*{-10pt}
  \includegraphics[width=3.4in]{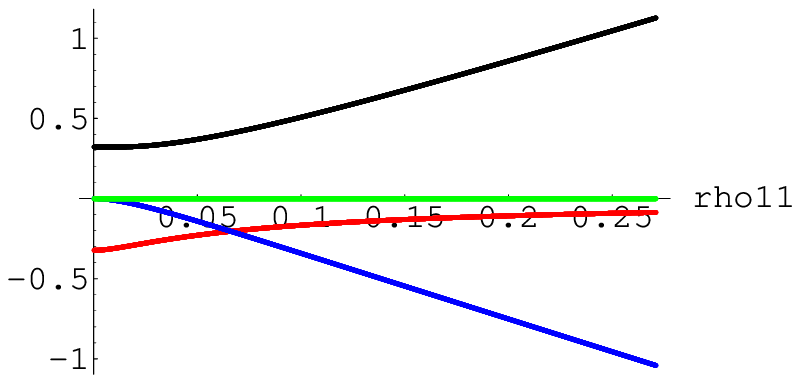}%
%  \hspace*{-20pt}
%  \vspace*{-10pt}
  \caption{
The left plot represents $a_h(\xi)$ for the supersymmetric mass deformation.
The red/blue/black points on the right plot represent $\a_h'(\xi)$, $2(\r_{10}-\xi \r'_{10}(\xi))$
and $-(\c_{10}(\xi)-\xi\c_{10}'(\xi))$ correspondingly. The green points on the right plot 
represent numerical verification of the first law of the thermodynamics, see \eqref{1sts}.
}
\label{figs2}
\end{figure}

Since the Sturm-Liouville problem is technically more complex here, we restrict our discussion to 
$\xi\ge 0$, or following \eqref{match2} $m^2\ge 0$. Results of the numerical analysis are presented in 
Figs.~\eqref{figs1}-\eqref{figs2}. Fitting the first 200 points of $\r_{10}(\xi)$ and $\c_{10}(\xi)$
with a 10th order polynomial in $\xi$ we find 
\begin{equation}
\r'_{10}(\xi)\bigg|_{\xi=0}^{fit}=-2.069(2)\,,\qquad \c_{10}\bigg|_{\xi=0}^{fit}=-0.3219(7)\,,
\eqlabel{fits}
\end{equation}
which should be compared with the high temperature predictions\footnote{We reduce the accuracy compare to the 
bosonic mass deformation case in order to speed-up computations. For a selected set of $41$ sets 
\eqref{susy4} we improve accuracy by roughly a factor of $10^2$. More accurate data does not provide 
noticeable quantitative difference.} \eqref{rel6}
\begin{equation}
\begin{split}
\r'_{10}(\xi)\bigg|_{\xi=0}^{high\ temperature}&=-3\ln 2\approx -2.07944\,,\\ 
\c_{10}(\xi)\bigg|_{\xi=0}^{high\ temperature}&=-\frac{2^{1/2}\Gamma\left(\frac 34\right)^4}{\pi^2}\approx 
-0.3231(1)\,.
\end{split}
\eqlabel{predsusy}
\end{equation}
We do not find a signature for the existence of $\r_{crit}$ --- it appears for supersymmetric 
deformation the boundary value problem \eqref{susy1} always have a solution\footnote{Recall for a bosonic 
deformation $\r_{crit}\approx -0.0358(2)$ which is well inside the set of $\xi$ considered in Fig.~\ref{figs1}.}.

Fig.~\ref{figs2} represent $a_h(\xi)$ (the left plot) and a numerical verification of the first law of thermodynamics 
\eqref{1sts} (the right plot):
\begin{equation}
\begin{split}
\biggl\{0\biggr\}_{green}=&\biggl\{2\left(\r_{10}(\xi)-\xi\frac{d\r_{10}(\xi)}{d\xi}\right)\biggr\}_{blue}+
\biggl\{-\left(\c_{10}(\xi)-\xi\frac{d\c_{10}(\xi)}{d\xi}\right)\biggr\}_{black}\\
&+\biggl\{\frac{da_h(\xi)}{d\xi}
\biggr\}_{red}\,,
\end{split}
\eqlabel{1stscolor}
\end{equation}
where subscripts refer to the color of the points on the right plot in Fig.~\ref{figs2}.
We find that \eqref{1stscolor} holds with an accuracy of $10^{-3}$.

\subsection{Generic deformation}
Though in this paper we study two special cases of the thermal PW flows --- one corresponding to 
 a bosonic mass deformation where the supergravity scalar $\c(x)\equiv 0$, and the other one 
corresponding to a supersymmetric mass deformation where the ratio of the coefficients
of the non-normalizable modes of the supergravity scalars $\{\r(x),\c(x)\}$ is the same as for the 
supersymmetric PW flow  --- the numerical methods developed here can be applied for a generic 
mass deformation as well. In fact, the Sturm-Liouville problem one has to solve in a generic case is exactly the same 
as in the supersymmetric case where the  boundary conditions are given by \eqref{susy1}. The only difference from the  
supersymmetric mass deformation case is the (fixed) relation between the  coefficients of the non-normalizable modes of the 
supergravity scalars $\{\r(x),\c(x)\}$ given by \eqref{not1a}
$$
\eta^2=6\ \frac{m_f^2}{m_b^2}\ \xi\,.
$$

Recall (see Fig.\ref{figb1}) that when $m_f=0$ there is a critical value $\r_{crit}$, such that for 
$\xi> \r_{crit}$ there is no nonsingular thermal PW flow. On the other hand (see Fig.\ref{figs1}), we do not find the  evidence for 
$\r_{crit}$ for the supersymmetric mass deformation $m_f=m_b$. Thus we expect that 
\begin{equation}
\r_{crit}\equiv \r_{crit}\left(\Delta\equiv \frac{m_f^2}{m_b^2}\right)\,,
\eqlabel{rcritdep} 
\end{equation}
such that $\r_{crit}\to \infty$ as $\frac{m_f^2}{m_b^2}\to \Delta^*\le 1$. 
As we show in subsection \ref{subsec:tpwb} the existence of $\r_{crit}$ for $m_f=0$ implies destabilization of the 
thermal plasma at low temperatures. What we appear to be finding is that this particular instability 
is cured by turning on $m_f\ne 0$. 
Verifying the latter conjecture is beyond the scope of this paper.  However, we would like to point out a different type of instability 
in a system closely related to the one studies here which is eliminated by  turning on sufficiently large $m_f$. 
Consider thermal PW flow with the flat $R^3$ in \eqref{ab} compactified on a three torus\footnote{Since a 
three torus is locally the same as $R^3$,
such a compactification does not change any of the above computations.} $T^3$. 
This represents holographic dual to thermal $\caln=2^*$ plasma on finite volume $T^3$. 
Analysis identical to that in \cite{bninst} shows  that this system is generically non-perturbatively unstable 
with respect to a spontaneous creation of the $D3\overline{D3}$ brane pairs\footnote{Similar instability 
was first identified in \cite{mm}.}. This instability is suppressed provided\footnote{See 
eq.(4.16) of \cite{bninst}
where one has to remove $(-2)$ from the inequality (coming from the curvature of the 3-space) and replace $\r_{11}$
of \cite{bninst} with $4\r_{11}$ due to a different choice of a radial coordinate.} 
\begin{equation}
\begin{split}
\c_0^2\ge 6\r_{11}\ge 0\qquad \Longleftrightarrow\qquad \eta^2\ge 6\xi\ge 0\qquad \Rightarrow\qquad  \Delta\equiv \frac{m_f^2}
{m_b^2}\ge 1     \,.
\end{split}
\eqlabel{qn1}
\end{equation}
So, for the nonperturbative instability due to $D3\overline{D3}$ spontaneous brane pair creation for the thermal PW flow  
compactified on $T^3$ we find that $\Delta^*=1$.

\subsection{Validity of the supergravity approximation}
\label{subsec:vsa}

\begin{figure}[t]
  \hspace*{-20pt}
\psfrag{rho11}{\raisebox{1ex}{\footnotesize\hspace{-0.8cm}$\qquad \xi$}}
\psfrag{k}{\raisebox{0.0ex}{$\k(\xi)$}}
\psfrag{rc}{\raisebox{0.5ex}{$\r_{crit}$}}
  \includegraphics[width=3.4in]{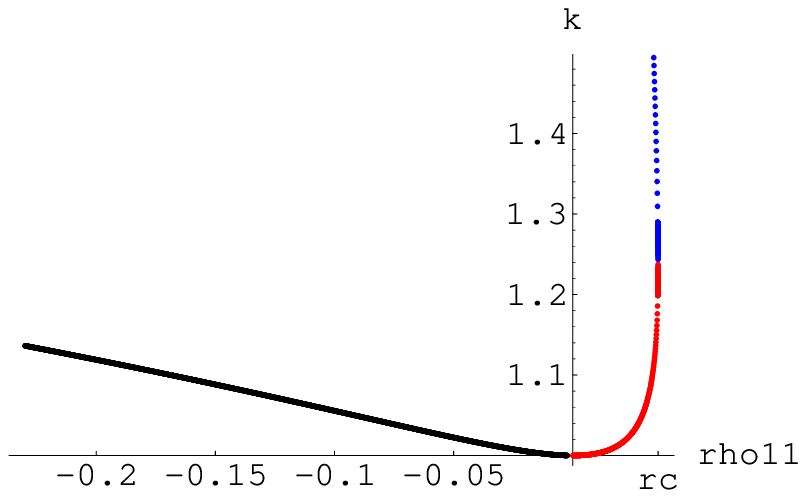}
%  \hspace*{-10pt}
  \includegraphics[width=3.4in]{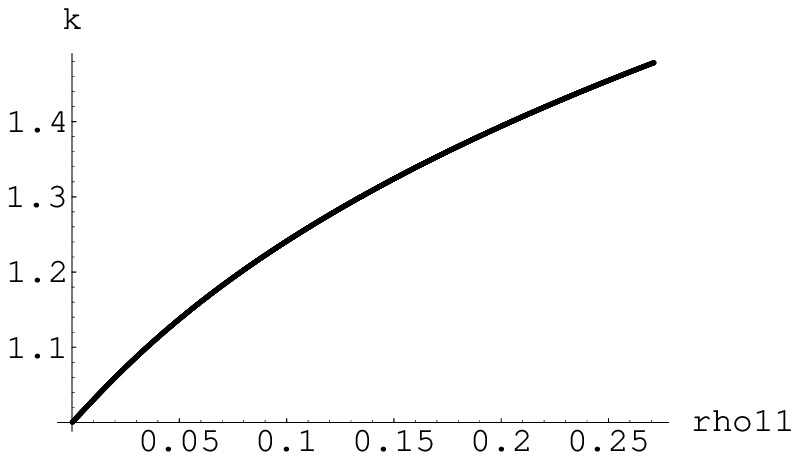}%
%  \hspace*{-20pt}
%  \vspace*{-10pt}
  \caption{
The left plot represents ratio of thermal PW geometry horizon curvature to 
its curvature at the boundary for bosonic mass deformation. The right plot represents 
this ratio for the supersymmetric mass deformation. 
}
\label{figcuv}
\end{figure}

Our analysis of the holographic dual to thermal $\caln=2^*$ gauge theory plasmas are done in the supergravity 
approximation to type IIb string theory. In the planar limit considered here the string loop corrections
are suppressed and the supergravity approximation  is valid provided curvature invariants of the thermal 
PW flow geometry \eqref{ab} are small. We find that along the flow from the boundary to the horizon
the absolute value of the Ricci scalar 
of the five dimensional thermal PW geometry \eqref{ab} increases. As a quantitative criteria for such 
an increase we propose the ratio of the scalar curvature at the horizon to the scalar curvature near the boundary
\begin{equation}
\kappa\equiv\frac{R^{horizon}}{R^{boundary}}=\frac{\calp^{horizon}}{\calp^{boundary}}=-\frac 43\ \calp^{horizon}\,,
\eqlabel{valid} 
\end{equation}
where we used \eqref{ee} and the fact that $\calp^{boundary}=-\frac 34$.
Values\footnote{In order to evaluate $\calp^{horizon}$ besides $a_h(\xi)$ we retained 
horizon values of the supergravity scalars $\{\r(x),\c(x)\}$ while solving the Sturm-Liouville problem 
\eqref{bose1} or \eqref{susy1}. } of $\kappa(\xi)$ for both bosonic and supersymmetric mass deformations are presented in Fig.~\ref{figcuv}.
Notice that for the 'blue' phase 
\begin{equation}
\lim_{\xi\to 0+}\k^{bosonic}_{blue}(\xi)=+\infty\,,
\eqlabel{sugrabreak}
\end{equation}
and thus supergravity approximation for the blue phase of the bosonic mass deformation for small 
values of $\xi$ is not valid.

%%%%%%%%%%%%%%%%%%%%%%%%%%%%%%%%%%%%%%%%
\section{$\caln=2^*$ plasma at finite temperature}
\label{sec:r}

In this section we discuss thermodynamics of the $\caln=2^*$ gauge theory plasma for two special cases:
\nxt $bosonic$ deformation, when $m_f=0$ and $m_b\ne 0$;
\nxt $supersymmetric$ deformation, when $m_f=m_b=m$.

\subsection{Thermal PW flows with $m_f=0$, $m_b\ne 0$}
\label{subsec:tpwb}

\begin{figure}[t]
  \hspace*{-20pt}
\psfrag{mbT2}{\raisebox{2ex}{\footnotesize\hspace{-0.5cm}$\qquad  \mu\equiv \frac{m_b^2}{T^2}$}}
\psfrag{mc2}{\raisebox{0.0ex}{$\mu_{crit}$}}
\psfrag{Fred}{\raisebox{0.5ex}{$\calf_{bosonic}/\left(\frac 18 \pi^2 N^2 T^4\right)\ \qquad\ $}}
\psfrag{vsb2}{\raisebox{0.5ex}{$3\ \times\ c_{s,bosonic}^2$}}
  \includegraphics[width=3.4in]{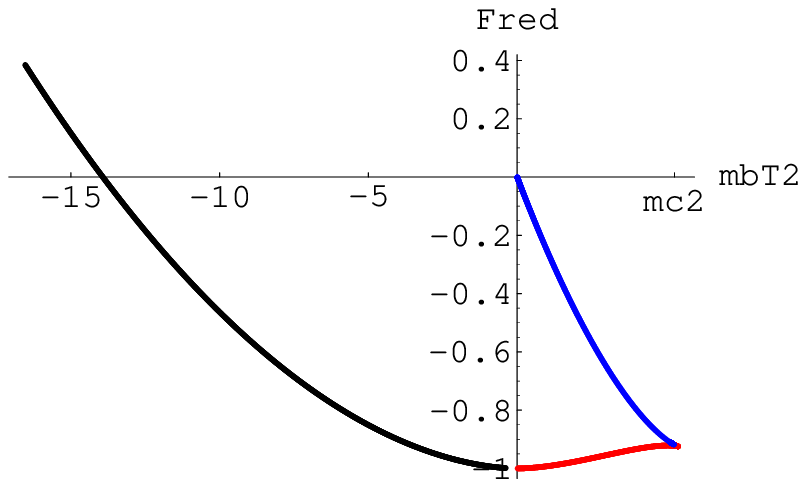}
%  \hspace*{-10pt}
 \includegraphics[width=3.4in]{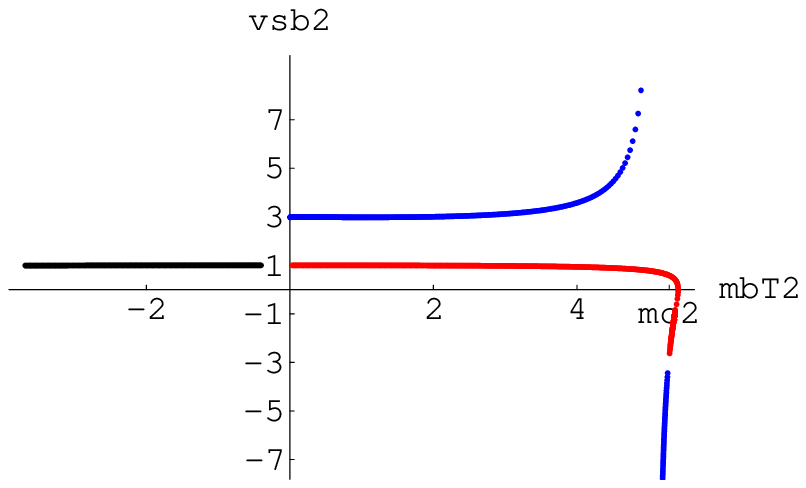}%
%  \hspace*{-20pt}
%  \vspace*{-10pt}
  \caption{
The left plot represents bosonic mass deformation free energy $\calf_{bosonic}$ as 
a function of $\mu\equiv \frac{m_b^2}{T^2}$. The critical mass is at $\mu_{crit}\approx 5.28(5)$. 
The right plot represents speed of sound squared $c_{s,bosonic}^2$ for 
bosonic mass deformation as a function of $\mu$.   Recall that the speed of sound in the conformal plasma 
$c_{s,CFT}^2=\frac 13$.
}
\label{figfreeb}
\end{figure}

\begin{figure}[t]
  \hspace*{80pt}
\psfrag{mbT2}{\raisebox{2ex}{\footnotesize\hspace{-0.5cm}$\qquad  \mu\equiv \frac{m_b^2}{T^2}$}}
\psfrag{mc2}{\raisebox{0.0ex}{$\mu_{crit}$}}
\psfrag{ms2}{\raisebox{0.0ex}{$\ \ \mu^*$}}
\psfrag{Fred}{\raisebox{0.5ex}{$\calf_{bosonic}/\left(\frac 18 \pi^2 N^2 T^4\right)\ \qquad\ $}}
\psfrag{vsb2}{\raisebox{0.5ex}{$3\ \times\ c_{s,bosonic}^2$}}
%  \includegraphics[width=3.4in]{freeb}
%  \hspace*{-10pt}
 \includegraphics[width=3.4in]{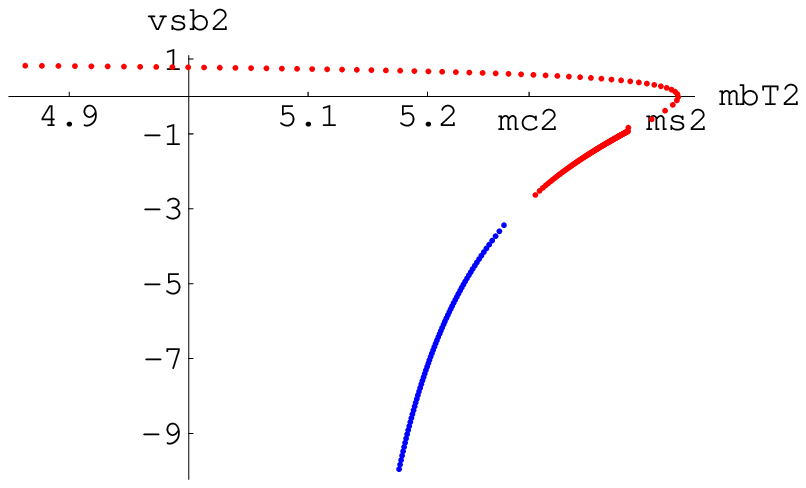}%
%  \hspace*{-20pt}
%  \vspace*{-10pt}
  \caption{
More detailed plot of the speed of sound squared $c_{s,bosonic}^2$ for 
bosonic mass deformation as a function of $\mu$. The speed of sound vanishes at $\mu^*\approx 5.40(9)$.
Notice that though $\mu^*>\mu_{crit}$, corresponding to $\mu^*$ value of $\xi^*\approx 0.035(1)$ is less than 
$\r_{crit}\approx 0.03582(2)$,
see Fig.~\ref{figb1}.    Recall that the speed of sound in the conformal plasma $c_{s,CFT}^2=\frac 13$.
}
\label{figsoundb}
\end{figure}

As explained in subsection \ref{subsec:bd}, for the bosonic deformation we observe two phases (`red' phase and 
`blue' phase) of the 
thermal gauge theory plasma for mass deformation parameter $\xi \in (0,\r_{crit})$ 
characterized by two different values for the bosonic condensate $\r_{10}(\xi)$. 
These are the red and the blue sets of points in Fig.~\ref{figb1}.
For large values of the bosonic deformation parameter ($\xi>\r_{crit}$ in Fig.~\ref{figb1})
thermal $\caln=2^*$ gauge theory plasma with bosonic deformation discussed does not exist.  
The set of black points in Fig.~\ref{figb1} corresponds to $\xi<0$, which following \eqref{match3}
corresponds to $m_b^2<0$. As we argued in subsection \ref{subsec:vsa}, we can trust supergravity approximation for 
the red set of points (everywhere) and for black points over the range we studies (corresponding to $m_b^2/T^2<-15$).
The supergravity approximation for the blue set of point is much less reliable\footnote{The smallest absolute value 
horizon curvature for the blue points is the same as the largest horizon curvature for the red 
points.}, especially for small values of $\xi$.

We use \eqref{match3}, \eqref{vs2}, \eqref{mff} and \eqref{speedb} to evaluate the free energy density and the speed of sound 
for the bosonic mass deformation. For the free energy density we set an overall (arbitrary) constant part of the free 
energy\footnote{Such ambiguity is present both in thermal quantum field theory and in the framework of the 
holographic renormalization of the string theory duals to thermal gauge theories.} 
\begin{equation}
\calf_{0}^{bosonic}=0\,.
\eqlabel{f00}   
\end{equation}
Notice that while in the supersymmetric case one can determine $\calf_{0}^{susy}$ unambiguously (see \eqref{fosusy})
by requiring the $\calf_{susy}(T=0)=0$, we can not do so for the bosonic deformation: as we describe shortly, 
the bosonic deformation becomes unstable both thermodynamically and dynamically at low temperatures.
Results of the analysis are presented in Figs.~\ref{figfreeb}-\ref{figsoundb}.

The left plot in Fig.~\ref{figfreeb} describes the free energy density for the bosonic deformation $\calf_{bosonic}$
relative (up to a sign) to the free energy density of the conformal gauge theory plasma
\begin{equation}
\calf_{CFT}=-\frac 18\ \pi^2 N^2 T^4\,,
\eqlabel{fcft}
\end{equation} 
as a function of 
\begin{equation}
\mu\equiv \frac{m_b^2}{T^2}\,.
\eqlabel{defmu}
\end{equation}
Notice that the blue phase has always a larger (or equal) free energy than the red phase
\begin{equation}
\calf_{bosonic,red}(\mu)\le \calf_{bosonic,blue}(\mu)\,,
\end{equation}
with equality achieved only at $\mu=\mu_{crit}$ corresponding to $\r_{crit}$ according to \eqref{match3}, \eqref{defmu}.
 For the red phase
\begin{equation}
\frac{\calf_{bosonic,red}(\mu=0)}{\ft 18\pi^2 N^2 T^4}=-1\,,
\eqlabel{red0}  
\end{equation}
in agreement with the fact that for $m_b=0$ (or at very high temperatures $T\to \infty$)
the free energy density must be that of the conformal gauge theory. 

The right plot in Fig.~\ref{figfreeb} and a more detailed plot in Fig.~\ref{figsoundb}
present the speed of sound squared for the thermal $\caln=2^*$ gauge theory plasma, relative to the speed of sound 
in the conformal gauge theory plasma
\begin{equation}
c_{s,CFT}^2=\frac 13\,,
\eqlabel{cscft}
\end{equation}
for bosonic deformation as a function of $\mu$.
Although bosonic deformation phase discussed here does not exist for $\xi>\r_{crit}$, 
bosonic phase actually extends to values of $\mu_{crit}<\mu \le \mu^*$ (see Fig.~\ref{figsoundb}). 
The reason for this is that relation between $\xi$ and $\mu$ as given by \eqref{match3} involves 
a factor $e^{6a_h(\xi)}$ and is not monotonic around  $\xi\sim \r_{crit}$. A value of 
$\xi^*$ corresponding to $\mu^*$ is actually less than $\r_{crit}$: 
\begin{equation}
\mu^* \approx 5.40(9)\ >\ \mu_{crit}\approx 5.28(5)\qquad \Rightarrow\qquad \xi^*\approx 0.035(1)\ <\ \r_{crit}\approx 0.03582(2) \,.
\eqlabel{msv}
\end{equation}
We find that the speed of sound for the blue phase is superlunimal, which is yet another indication that this 
phase is unphysical. The speed of sound diverges for the blue phase as one approach $\mu_{crit}$
either from above or below. The speed of sound for the red phase at $\mu=0$ is that as in the CFT plasma;
it vanishes at $\mu=\mu^*$ and becomes purely imaginary for $\xi^*<\xi< \r_{crit}$ (still for the red phase
and well within supergravity approximation) 
\begin{equation}
\begin{split}
&c_{s,bosonic,red}^2(\mu=0)=\frac 13\,,\qquad c_{s,bosonic,red}^2(\mu=\mu^*)=0\,,\\
&c_{s,bosonic,red}^2\left(\mu\equiv 12\sqrt{2}\pi^2e^{6a_h(\xi)}\xi\right)< 0\,,\qquad \xi^*<\xi<\r_{crit}\,.
\end{split}
\eqlabel{csbos}
\end{equation}
We would like to stress that conclusions \eqref{csbos} are very robust and in fact, immediately follow from the 
existence of $\r_{crit}$ for the bosonic condensate $\r_{10}(\xi)$. 
Indeed, existence of $\r_{crit}$, \ie,\ impossibility to determine $\r_{10}(\xi)$ for $\xi>\r_{crit}$,
along with the assumption for the smoothness of $\r_{10}(\xi)$, implies that $\rho_{10}(\xi)$ is not a single valued 
function of $\xi$ around $\r_{crit}$; moreover, 
\begin{equation}
\lim_{\xi\to \r_{crit}-0}\r_{10}'(\xi)=\mp \infty\,,
\eqlabel{bin1}
\end{equation}
with the minus sign in the red phase and the plus sign in the blue phase (precisely what is presented in Fig.~\ref{figb1}).
But then from \eqref{speedb}
\begin{equation}
\lim_{\xi\to \r_{crit}-0}\Sigma_{bosonic}(\xi)=\mp \infty\,,
\eqlabel{bin2}
\end{equation}
again with the minus sign in the red phase and the plus sign in the blue phase.
Thus using \eqref{vs2} we see that in the red phase 
\begin{equation}
\lim_{\xi\to \r_{crit}-0}\biggl(3\ \times\ c_{s,bosonic}^2(\xi)\biggr)=-3\,,
\eqlabel{bin3}
\end{equation}
(in agreement with Figs.~\ref{figfreeb}-\ref{figsoundb}), and for the blue phase 
\begin{equation}
\lim_{\xi\to \xi_{\infty}\mp 0}\biggl(3\ \times\ c_{s,bosonic}^2(\xi)\biggr)=\pm \infty\,,\qquad \Sigma_{bosonic,blue}(\xi_{\infty})=3\,,
\eqlabel{bin4}
\end{equation}
where $0<\xi_{\infty}<\r_{crit}$.  Since in the red phase at $\xi=0$ the speed of sound squared is $\frac 13>0$
and given \eqref{bin3} there is a guaranteed value of $\xi^*<\r_{crit}$ at which the speed of sound vanishes, 
and for $\xi^*<\xi<\r_{crit}$ the speed of sound is purely imaginary. 
As emphasized in \cite{binst}, 
a thermal system with $c_s^2<0$ is unstable simultaneously thermodynamically (it has a negative specific heat)
and dynamically (amplitude of small pressure/energy density fluctuations exponentially grows).  

Finally, we do not see thermodynamic instability in thermal $\caln=2^*$ plasma with bosonic mass deformation for 
$m_b^2<0$, see  Fig.~\ref{figfreeb}. In fact, we find that for negative values of $\mu$ the speed of sound 
slowly (and monotonically) decreases --- it reaches about $90\%$ of the conformal plasma speed of sound \eqref{cscft}
at\footnote{This is outside the range of the right plot in Fig.~\ref{figfreeb}.} $\mu\approx -35$. 
Of course, this does not exclude the fact that thermal $\caln=2^*$ plasma with $m_b^2<0$ might have dynamical instability
without  accompanying thermodynamics instability \cite{binst}.

\subsection{Thermal PW flows with $m_f=m_b\equiv m$}

\begin{figure}[t]
  \hspace*{-20pt}
\psfrag{mT}{\raisebox{2ex}{\footnotesize\hspace{-1.0cm}$\qquad \frac{m}{T}$}}
\psfrag{Fs}{\raisebox{0.5ex}{$f_{susy}\equiv -\calf_{susy}/\left(\frac 18 \pi^2 N^2 T^4\right)\ \qquad\ $}}
\psfrag{lnf}{\raisebox{0.5ex}{$\ln\left(f_{susy}\right)$}}
  \includegraphics[width=3.4in]{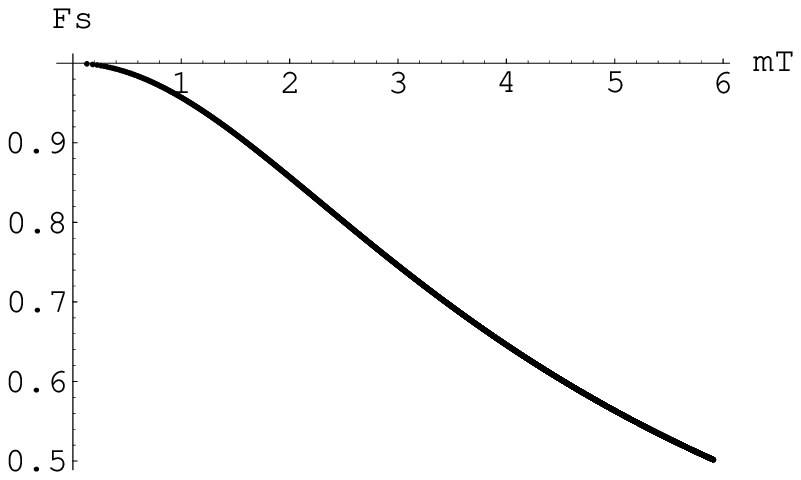}
%  \hspace*{-10pt}
 \includegraphics[width=3.4in]{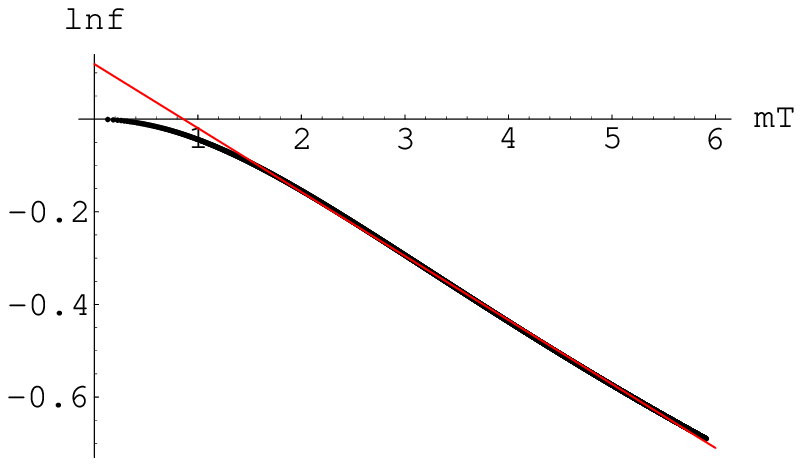}%
%  \hspace*{-20pt}
%  \vspace*{-10pt}
  \caption{
The left plot represents supersymmetric mass deformation free energy $\calf_{susy}$ as 
a function of $\frac{m}{T}$. The right plot represents $\ln(f_{susy})$ as a function of 
$\frac{m}{T}$ (black points) and a straight line fit (red curve) for the last 2500 points
of the distribution.
} 
\label{figfrees}
\end{figure}

\begin{figure}[t]
  \hspace*{80pt}
\psfrag{mT}{\raisebox{2ex}{\footnotesize\hspace{-0.5cm}$\frac{m}{T}$}}
\psfrag{vs2}{\raisebox{0.5ex}{$3\ \times\ c_{s,susy}^2$}}
%  \includegraphics[width=3.4in]{freeb}
%  \hspace*{-10pt}
 \includegraphics[width=3.4in]{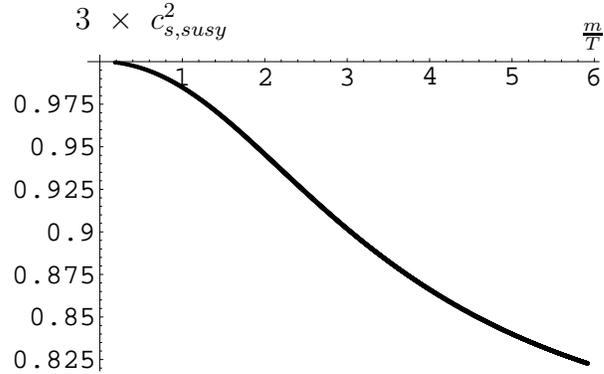}%
%  \hspace*{-20pt}
%  \vspace*{-10pt}
  \caption{
The speed of sound squared $c_{s,susy}^2$ for 
supersymmetric mass deformation as a function of $\frac{m}{T}$.
}
\label{figsounds}
\end{figure}

Bosonic and fermionic condensates $\{\r_{10}(\xi), \chi_{10}(\xi)\}$ 
for thermal $\caln=2^*$ gauge theory plasma with supersymmetric mass deformation are presented in Fig.~\ref{figs1}. 
As we argued in subsection \ref{subsec:vsa}, supergravity approximation here is valid of the whole range of deformation 
parameter $\xi$ considered. We use \eqref{match3}, \eqref{vs2}, \eqref{mfs}, \eqref{speeds} and \eqref{fosusy} 
to evaluate the free energy density and the speed of sound for the supersymmetric mass deformations.  
Results of the analysis are presented in Figs.~\ref{figfrees}-\ref{figsounds}.
We do not see any signature of the phase transition\footnote{This would show up as the free energy 
$\calf_{susy}\left(\frac mT\right)$ changing sign for a particular 
value of $\frac mT$. }, or the thermodynamic instability. 

Probably the most unexpected result of the analysis is that strongly coupled thermal $\caln=2^*$ plasma 
for supersymmetric mass deformation behaves much closer to a conformal plasma than one would naively expect. 
The right plot on Fig.~\ref{figfrees} presents the logarithm of the ratio  of the $\caln=2^*$ plasma free energy density
to the free energy density of the conformal plasma \eqref{fcft}  
\begin{equation}
\ln(f_{susy})\equiv \ln \frac{\calf_{susy}}{\calf_{CFT}} \,,
\eqlabel{fsusydef}
\end{equation}
as a function of $\frac mT$. 
We did a straight line line fit (solid red line on the right plot on Fig.~\ref{figfrees}) for the low temperature\footnote{In practice, 
we define low temperatures as the subset of  the mass deformation parameters $\frac mT>\sim 1$.} 
\begin{equation}
\ln(f_{susy})\bigg|^{FIT}\left(x\right)=0.118(8) - 0.138(1)\ x\,.
\eqlabel{fitf}
\end{equation}
Thus, to a good approximation the free energy of the thermal $\caln=2^*$ plasma can be approximated by 
\begin{equation}
\calf_{susy}^{approx}\ \approx\ 1.1(3)\times \calf_{CFT}\ e^{-\frac{m}{7.2(4)\ T}}\ \propto\ - T^4\ e^{-\frac{m}{7\ T}}\,,
\eqlabel{fapprox}
\end{equation}
so that noticeable deviation from conformality occurs at $T\sim \frac 17 m$.

It is instructing to compare \eqref{fapprox} with the zero coupling result for the $\caln=2^*$ plasma.
At zero coupling we have\footnote{We neglect $(-1)$ in the large $N$ limit.} $N^2$ massless $\caln=2$  vector multiples and $N^2$ massive $\caln=2$ 
hypermultiplets. The free energy density of the massless states is just $\frac 23 \calf_{CFT}$ in \eqref{fcft}, while the 
free energy density of a mass $m$ spin $s$ particle is given by \cite{dj,a} 
\begin{equation}
\calf^{s}=-\frac{(2s+1)m^2T^2}{2\pi^2}\sum_{\ell=1}^{\infty} \frac{\eta^{\ell+1}}{\ell^2} K_2\left(\frac{\ell m}{T}\right)\,,
\eqlabel{frees}
\end{equation}
where $\eta=\pm 1$ for bosons/fermions correspondingly. As low temperatures $\frac mT\gg 1$ we can use asymptotic expansion 
for the Bessel function $K_2(x)$ 
\begin{equation}
K_2(x)\approx \sqrt{\frac{\pi}{2x}}\ e^{-x}\,,\qquad x\gg 1\,,
\eqlabel{k2ex}
\end{equation}
to conclude that 
\begin{equation}
\calf^{s}\approx -\frac{(2s+1)m^{3/2}T^{5/2}}{(2\pi)^{3/2}}\ e^{-\frac{m}{T}}\,,\qquad \frac{m}{T}\gg 1\,.
\eqlabel{freea}
\end{equation}
Altogether we find  that the free energy of the thermal $\caln=2^*$ plasma for supersymmetric mass deformation at zero coupling is 
\begin{equation}
\calf_{\caln=2^*}=-\frac{\pi^2 N^2T^4}{12}\left(1+\frac{24\sqrt{2}}{\pi^{7/2}}\ \left(\frac mT\right)^{3/2}\ e^{-\frac mT}\right)\,,
\qquad  \frac mT\gg 1\,.
\eqlabel{free0c}
\end{equation}
From \eqref{free0c}, we find that the free energy density is roughly $60\%$ of the high temperature result at $T\sim m$, 
while the corresponding number for the strongly coupled plasma \eqref{fitf} is about $98\%$.

The hydrodynamic properties and the jet quenching of strongly coupled nonconformal gauge theory plasma are 
typically parameterized in terms of \cite{bbs,bjet} $\delta$
\begin{equation}
\delta\equiv \left(\frac 13-c_s^2\right) \,.
\end{equation}
For the strongly coupled $\caln=2^*$ thermal plasma at\footnote{The temperature of the 
quark-gluon plasma at RHIC is of order the QCD deconfinement temperature \cite{shu}.} $T\sim 1.5 m$ we find (see Fig.~\ref{figsounds}) 
\begin{equation}
\delta^{\caln=2^*}|_{T\approx 1.5 m}\approx 2\times 10^{-3}\,,
\eqlabel{delta}
\end{equation}
which further emphasizes the striking similarity between strongly coupled $\caln=2^*$ plasma and 
conformal $\caln=4$ gauge theory plasma.

%%%%%%%%%%%%%%%%%%%%%%%%%%%%%%%%%%%%%%%%
\section{Conclusion and future directions}
\label{sec:cfd}

\begin{figure}[t]
  \hspace*{-20pt}
\psfrag{mT}{\raisebox{2ex}{\footnotesize\hspace{-1.0cm}$\qquad \frac{m}{T}$}}
\psfrag{mbT}{\raisebox{2ex}{\footnotesize\hspace{-1.0cm}$\qquad \frac{m_b}{T}$}}
\psfrag{es}{\raisebox{0.5ex}{$\cale_{susy}/\left(\frac 38 \pi^2 N^2 T^4\right)\ \qquad\ $}}
\psfrag{eb}{\raisebox{0.5ex}{$\cale_{bosonic}/\left(\frac 38 \pi^2 N^2 T^4\right)\ \qquad\ $}}
\psfrag{ms}{\raisebox{0ex}{\footnotesize\hspace{0.0cm}$\sqrt{\mu^*}$}}
 \includegraphics[width=3.4in]{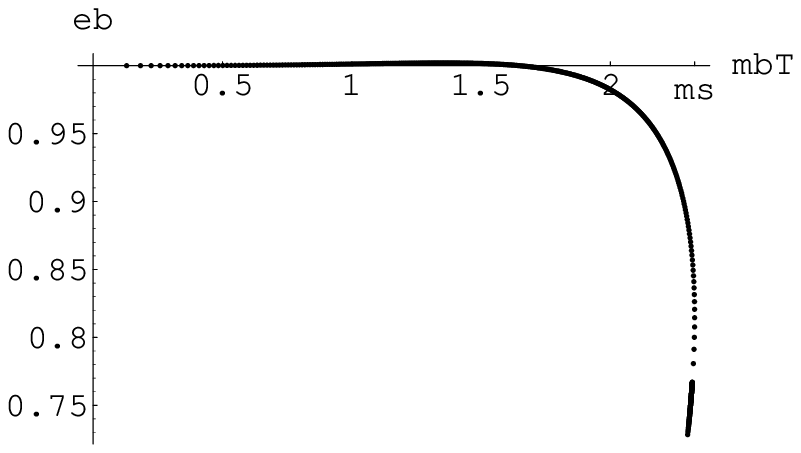}
%  \hspace*{-10pt}
 \includegraphics[width=3.4in]{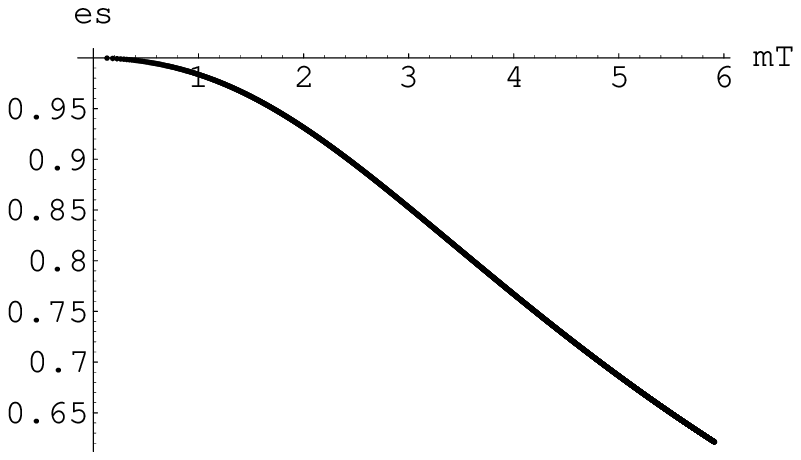}%
%  \hspace*{-20pt}
%  \vspace*{-10pt}
  \caption{
The left plot represents bosonic mass deformation energy density $\cale_{bosonic}$ as 
a function of $\frac{m_b}{T}$. The value $\sqrt{\mu^*}$  (see Fig.~\ref{figsoundb}) denotes the ratio $\frac{m_b}{T}$
for which the speed of sound vanishes.
The right plot represents supersymmetric mass deformation energy density $\cale_{susy}$ as 
a function of $\frac{m}{T}$.
} 
\label{figenergy}
\end{figure}

In this paper we presented detailed analysis of the thermodynamics of the strongly coupled $\caln=2^*$
gauge theory plasma. We considered two special cases of the mass deformations: bosonic mass deformation 
(where only the bosonic components of the $\caln=2$ hypermultiplet get nonzero mass) and supersymmetric  
mass deformation (with bosonic and fermionic components of the hypermultiplet getting the same mass).
We argued that both our theoretical framework and the numerical procedures are under control,
given a highly nontrivial consistency check of the first law of thermodynamics for the extracted data.
We also argued that the supergravity approximation is under control, as the curvature invariants of the 
dual geometry remain small. 

The most important result of the analysis is the striking similarity between thermodynamics of the 
mass deformed $\caln=2^*$ gauge theory plasma at temperatures of order the mass scale  
and the thermodynamics of the conformal $\caln=4$ SYM plasma. We would like to reemphasize the latter fact by presenting the plots 
for the energy energy for various mass deformations, see Fig.~\ref{figenergy}.

We could like to concluded with some open problems.
\nxt
In is important to generalize our analysis to other examples of the strongly coupled nonconformal gauge theory 
plasma. Most notably the cascading gauge theory plasma \cite{kt1,kt2,kt3,aby}.
\nxt It is interesting to evaluate hydrodynamic properties and jet quenching in $\caln=2^*$ gauge theory 
plasma in the low temperature regime. 
\nxt The speed of sound waves in $\caln=2^*$ gauge theory plasma for the bosonic mass deformation was 
found to vanish for certain value of $\frac{m_b}{T}$. It is interesting to study 
hydrodynamics of the gauge theory plasma in the regime of the vanishing speed of sound and determine 
the critical exponent of the bulk viscosity.

%%%%%%%%%%%%%%%%%%%%%%%%%%%%%%%%%%%%%%%%
\section*{Acknowledgments}

We would like to thank Jaume Gomis, Chris Herzog, Pavel Kovtun, Gerry McKeon
and Andrei Starinets for interesting discussions. 
We would like to thank Ofer Aharony and Rob Myers for valuable comments on the manuscript. 
AB's research at Perimeter
Institute is supported in part by the Government of Canada through
NSERC and by the Province of Ontario through MEDT.  AB gratefully
acknowledges further support by an NSERC Discovery grant.  JTL's research
was supported in part by the US Department of Energy under grant
DE-FG02-95ER40899 and the National Science Foundation under grant
PHY99-07949.  JTL acknowledges the hospitality of the KITP, where this
work was initiated.

%%%%%%%%%%%%%%%%%%%%%%%%%%%%%%%%%%%%%%%%

\end{document}